\documentclass[12pt]{JHEP3}

\usepackage{graphicx}
\usepackage{graphics}
\usepackage{epsfig}
\usepackage{ulem}
\usepackage{cite}

\title{The resummation of inter-jet energy flow for gaps-between-jets processes at HERA}

\author{R.~B.~Appleby and M.~H.~Seymour \\ Theory Group, Department of Physics and Astronomy, Schuster Laboratory,
\\ University of Manchester, Manchester, UK, M13 9PL \\
E-mail: \email{robert@theory.ph.man.ac.uk}, \email{seymour@theory.ph.man.ac.uk}}

\preprint{hep-ph/0308086\\MC-TH-2003-8}


\abstract{We calculate resummed perturbative predictions for gaps-between-jets processes and compare
to HERA data. Our calculation of this non-global observable 
needs to include the effects of primary gluon emission (global logarithms) and secondary gluon emission (non-global
logarithms)  to be correct at the leading logarithm (LL) level. We include primary emission by calculating anomalous dimension matrices 
for the geometry of the specific event definitions and estimate the effect of non-global logarithms in the 
large $N_c$ limit. The resulting predictions
for energy flow observables are consistent with experimental data.}

\begin{document}

\section{Introduction}

\label{secintro}

The subject of interjet energy flow \cite{Marchesini:1988} has attracted considerable interest ever since 
it was proposed \cite{dokshitzer,Bjorken:1992er} as a way to study 
rapidity gap processes using the tools of perturbation theory. Rapidity gap processes are defined as processes
containing two high $p_t$ jets with the region of rapidity between the jets containing nothing more than soft
radiation. This region is known generically as the gap. The presence of a range of scales offers a chance to study 
the interface between the soft, non-perturbative scales and the hard, perturbative scales of
QCD. 

In this paper we will calculate the perturbative contribution to gaps-between-jets cross sections in photoproduction 
at HERA, which have 
been measured by the ZEUS \cite{Derrick:1995pb,zeus:2003} and H1 
\cite{Adloff:2002em} 
collaborations.  A feature of the recent analyses is the use of a clustering algorithm to define the hadronic final state 
and hence the gap. 
The restriction of transverse radiation in a region of phase space, defined as $\Omega$ and directed away
from the observed jets and the beam directions, produces logarithms at each order of QCD perturbation theory of
the interjet energy flow, $Q_{\Omega}$, over some hard scale, $Q$. The precise definition of the restricted 
region, or gap, is totally free and in this work we are interested in the gap region defined by experimental 
rapidity gap analyses. The source of the large logarithms is twofold. The so-called
primary (or global) logarithms arise from radiation emitted directly into $\Omega$; these wide-angle gluons decouple 
from the dynamics of the colour-singlet jets and are described by an effective, eikonal theory 
\cite{Berger:2001ns,Oderda:1998en,Oderda:1999kr,Kidonakis:1998nf,Kidonakis:1998bk}.
 The second source of leading
logarithms arise from gluons emitted outside of the gap region, an area of phase space generically denoted
as $\bar{\Omega}$, which subsequently radiate into $\Omega$. These terms are known as non-global (secondary)
logarithms, or NGLs \cite{Dasgupta:2001sh,Dasgupta:2002bw,Appleby:2002ke,Appleby:2003ai}.

The primary logarithms are resummed using the formalism of Collins, Soper and Sterman (CSS)
\cite{Collins:ig,Sotiropoulos:1993rd,Contopanagos:1996nh,Kidonakis:1998bk}. In this method 
the cross section is factorised into a soft
part describing the emission of soft, wide angle gluons up  to scale $Q_{\Omega}$ and a hard part, 
describing harder quanta. 
A unique feature of QCD
is that the soft and the hard functions are expressed as matrices in the space of possible colour flow of the system.
The scale invariance and factorisation
properties of the cross section are then exploited to resum primary logarithms of $Q_{\Omega}/Q$. 
This resummation is driven by the ultraviolet pole parts of eikonal Feynman graphs and we write the resummed cross section 
in terms of the eigenvalues 
of $\Omega$-dependent soft anomalous dimension matrices. These matrices are known for gap definitions based on the
cone definition of the final state \cite{Oderda:1998en,Oderda:1999kr} and for a gap defined as a square patch in 
rapidity and azimuthal angle
\cite{Berger:2001ns};
here we are interested in gaps defined in terms of the clustering algorithms employed in the recent 
analyses. Hence we are 
required to calculate the corresponding anomalous dimension matrices.

The NGLs \cite{Dasgupta:2002bw,Dasgupta:2001sh} are unable to be incorporated into the resummation of the 
primary logarithms, because the gluon 
emission patterns that produce the NGLs are sensitive to underlying colour flows not included in the formalism. 
The effect of NGLs, which is a suppressive effect, on energy flow processes has been studied using numerical methods 
in the large~$N_c$ limit and overall factors describing their effect have been extracted for a two jet system, both
without \cite{Dasgupta:2002bw} and with \cite{Appleby:2002ke} clustering. This factor is not directly applicable to the 4 jet systems\footnote{Note that for a two-to-two process the incoming and outgoing partons radiate, so we 
consider the process to be of ``four jet'' type, although only two jets are seen in the final state.} relevant in the 
photoproduction of jets but, in the lack of a four jet formalism, we nevertheless use the two-jet factor in our predictions.

Our aim is to derive LL resummed predictions for the gap cross section, with primary logarithms correct
to all orders and secondary logarithms correct in the large~$N_c$ limit. The gap cross section will follow the HERA
analyses and demand two hard jets, defined using the kt clustering algorithm 
\cite{Catani:1993hr,Ellis:tq,Butterworth:2002xg}, and we will closely follow the
H1 and ZEUS gap definition. The technical aspects of soft gluon resummation
give a strong dependence on the gluon emission phase space, and hence a considerable part of our work will be 
concerned with the
calculation of soft gluon effects for the specific detector geometry of the H1 and ZEUS experiments.

The organisation of this paper is as follows. Section \ref{sechera} describes, in detail, 
the energy flow analyses of H1 and ZEUS. We 
describe the experimental cuts employed and the range of measured observables. We also discuss the theoretical
implementation of the inclusive kt algorithm employed to define the hadronic final state and the impact
on soft gluon resummation. Section \ref{secfact} describes the theoretical definition of our cross section and we employ the
standard QCD factorisation theorems to write it as the convolution of non-perturbative parton distributions and
a short-distance hard scattering function. We then proceed to refactorise the hard scattering function and exploit this 
factorisation to resum the large interjet logarithms. Section \ref{seckt} then derives the 
soft anomalous dimension matrices 
for the kt defined final state and in section \ref{secresults} we present detailed predictions of 
rapidity gap processes and compare 
to the H1 data. Finally we draw our conclusion in section \ref{secconc}. 
We find that our description of the data is good, although the approximate treatment of NGLs results in a
relatively large normalisation uncertainty.

\section{The HERA energy flow analyses}

\label{sechera}

In this section we will outline the experimental analyses of the photoproduction of gaps-between-jets processes and 
discuss the experimental cuts and rapidity gap observables. We will also describe the clustering algorithm used
to define the final hadronic state in the more recent ZEUS \cite{zeus:2003} and H1 \cite{Adloff:2002em} analyses.

The data for these events were collected when HERA collided $27.6$~GeV positrons\footnote{The positron energy 
varied a negligible amount between the two sets of analyses.}  with $820$~GeV protons, giving 
a centre of mass energy of $\sqrt{s}\simeq 300$~GeV. Following the jet-finding phase, which we will comment on
later, the total transverse energy flow between the two highest $E_T$ jets, denoted $E_T^{\mathrm{GAP}}$, is calculated
by summing the transverse energy of all particles that are not part of the dijets in the pseudorapidity region between
the two highest jets. An event is then defined as a gap event if the energy is less than some energy cut
$E_T^{\mathrm{CUT}}\equiv Q_{\Omega}$. A gap fraction is then calculated by dividing the cross section at fixed
$E_T^{\mathrm{CUT}}$ by
the inclusive cross section. The ZEUS collaboration performed a rapidity gap analysis 
several years ago~\cite{Derrick:1995pb} using the
cone algorithm for the jet definition and presented the gap fraction at $Q_{\Omega}=0.3$~GeV. We consider this 
value of $Q_{\Omega}$ as being too small for our perturbative analysis and will not make any predictions for this
data set. 
The more recent H1 and ZEUS analyses used the kt definition of the 
final state and both collaborations presented the gap fraction at four different values of $Q_{\Omega}$, as shown in 
table \ref{heracuts}. We will make predictions and compare to data for the  H1 data sets and, due 
to the fact that the ZEUS data is still preliminary, confine ourselves to making predictions for
the ZEUS analysis. We have summarised the cuts used in table \ref{heracuts}. 

\TABLE{
\begin{tabular}{|c|c|c|}
 \hline
 & H1 & ZEUS  \\ \hline \hline
$E_T^{\mathrm{jet1}}$ &  $>6.0$~GeV &  $>6.0$~GeV \\ \hline
$E_T^{\mathrm{jet2}}$ &  $> 5.0$~GeV &  $> 5.0$~GeV \\ \hline
$\eta^{\mathrm{jet1}}$ & $< 2.65$ &  $< 2.4$\\ \hline
$\eta^{\mathrm{jet2}}$ & $< 2.65$ & $< 2.4$\\ \hline
$\Delta\eta$ & $ 2.5 < \Delta\eta < 4$ & $2 < \Delta\eta <4$\\ \hline
$\eta_{\mathrm{jj}}$ & N/A & $< 0.75$ \\ \hline
$y$ & $0.3 < y < 0.6$ & $0.2 < y < 0.85$ \\ \hline
$Q^2$ & $< 0.01$~GeV$^2$ & $< 1$~GeV$^2$ \\ \hline
jet def. & kt & kt \\ \hline
gap def. & $\Delta y=\Delta\eta$ & $\Delta y=\Delta\eta$ \\ \hline
$R$ & $1.0$  & $1.0$ \\ \hline
$Q_{\Omega}$ & $0.5,1.0,1.5,2.0$~GeV  & $0.5,1.0,1.5,2.0$~GeV \\ \hline
\end{tabular}
\caption{The experimental cuts used for the HERA analyses.}
\label{heracuts}
}

\subsection{The \boldmath{kt} algorithm}

Of special interest to those going about soft gluon calculations is the method used to define the hadronic final state, 
the reason being that this jet-finding process determines the phase space
for soft gluon emission; the method used in the H1 and ZEUS data sets 
is the inclusive kt algorithm \cite{Catani:1993hr,Ellis:tq,Butterworth:2002xg}.
In this algorithm the final state is represented by a set of ``protojets'' $i$ with momenta $p_i^{\mu}$
 and works in an
iterative way, grouping pairs of protojets together to form new ones. The aim is to group almost-parallel protojets together
so that they are part of the same protojet. Once certain criteria are met, a protojet is considered a jet and is not considered
further. Here we follow the so-called inclusive scheme used at H1 and ZEUS which depends on the parameter R, normally set to unity. If we assume that any radiation into the gap is much softer than any parent radiation, then this radiation
 with $E_T < E_T^{\mathrm{jet}}$ will be merged into the jet (with kinematical variables 
$(\eta_{\mathrm{jet}},\phi_{\mathrm{jet}})$) if it
satisfies
\begin{equation}
(\eta_r-\eta_{\mathrm{jet}})^2+
(\phi_r-\phi_{\mathrm{jet}})^2 
< R^2,
\end{equation}
where we denote the kinematical variables of the radiated gluon by $(\eta_r,\phi_r)$.
Once merged, a gluon will be pulled out of the gap and can no longer produce a primary or secondary logarithm.
The gap is defined as the interjet region minus the
region of clustered radiation around the jets and may contain soft protojets. The gap transverse energy is then 
defined by the (scalar) sum of the protojets within the gap region, $\eta_1 < \eta < \eta_2$.

The kt gap definition can be contrasted to the older ZEUS analysis \cite{Derrick:1995pb}, which used the well known 
cone definition of the
final state with $R=1.0$. The gap transverse energy is then defined as the scalar sum of the hadrons within it, 
$\eta_1+R < \eta < \eta_2-R$.

\section{Factorisation, refactorisation and resummation of the cross section}

\label{secfact}

In this section we will exploit the standard factorisation theorems of QCD to write down the dijet production
cross section from the interaction of a proton and a positron. We will then refactorise the hard scattering function
into the product of two matrices in the space of possible hard scattering colour flow, one matrix describing
soft gluons radiated into the gap region and the other a hard scattering matrix. The renormalisation properties of the
cross section are then used to resum primary interjet logarithms, and write the result in terms of the
eigenvalues of the matrix of counterterms used to renormalise the soft function. In the following section we will 
calculate these matrices and their eigenvalues.

\subsection{Photoproduction cross sections}

The scattering of positrons and protons at HERA proceeds predominantly through the exchange of photons with
very small virtuality and produces a large subset of events with jets of high transverse momentum, $E_T$. 
The presence of this large scale allows the 
application of the perturbative methods of QCD to predict the cross section for multiple jet production.
This process is otherwise 
known as jet photoproduction.

The leading order (LO) QCD contribution can be divided into two types \cite{Oderda:1998en}. The first is 
the direct process in which the photon
interacts directly with a parton from the proton and proceeds through either the Compton process, $\gamma q \rightarrow gq$, or the photon-gluon fusion process, $\gamma g \rightarrow q\bar{q}$. The second contribution is the resolved contribution, in which
the virtual photon fluctuates into a hadronic state that acts as a source of partons, which then scatter off the partonic content of the proton. Therefore
the reaction proceeds through standard QCD $2\rightarrow2$ parton scattering processes. Note that the
precise determination of the partonic content of the photon is a very open question and there is a 
relatively large error associated with the photonic parton densities.
The spectrum of virtual
photons is approximated by the Weiz\"{a}cker--Williams \cite{williams:1934} formula,
\begin{equation}
F_{\gamma/e}(y)=\frac{\alpha}{2\pi}\frac{(1+(1-y)^2)}{y}\log\left( \frac{Q^2_{\mathrm{max}}(1-y)}{m_e^2 y^2}\right),
\end{equation}
where $m_e$ is the electron mass, $y$ is the fraction of the positron's energy that is transfered to the photon, and $Q^2_{\mathrm{max}}$ is the maximum
virtuality of the photon, which is determined by the experimental cuts employed in the analyses. 
Then, by using the equivalent photon approximation, the cross section for the process $e^+p\rightarrow e^{+}X$ is given by
the convolution
\begin{equation}
\mathrm{d}\sigma(e^+p\rightarrow e^{+}X) =\int_{y_{min}}^{y_{max}} \mathrm{d}y\,F_{\gamma/e}(y) 
\,\mathrm{d}\sigma(\gamma p \rightarrow X),
\end{equation}
where we write $\mathrm{d}\sigma(\gamma p \rightarrow X)$ for the cross section of $\gamma p \rightarrow X$. The
centre of mass energy squared for the photon-proton system is $W^2=ys$, where $s$ is the centre of mass energy squared for the positron-proton system. At HERA,
$s\simeq 90,000$~GeV$^2$ and the values for $y_{\mathrm{min}}$ and $y_{\mathrm{max}}$ are 
determined by the experimental analyses.
We can now write down the specific expression for the production of two high $E_T$ jets from the photon-proton system, which is written as a sum
of the direct and resolved contributions,
\begin{eqnarray}
d\sigma_{e^+p}(s,\hat{t},\Delta\eta,\alpha_s(\mu_r),Q_{\Omega})
=&& \int_{y_{min}}^{y_{max}} \mathrm{d}y\,F_{\gamma/e}(y) \bigg(
\mathrm{d}\sigma_{\gamma p}^{\mathrm{dir}}
(s_{\gamma p},\hat{t},\Delta\eta,\alpha_s(\mu_r),Q_{\Omega}) \nonumber \\
&&+ \mathrm{d}\sigma_{\gamma p}^{\mathrm{res}}
(s_{\gamma p},\hat{t},\Delta\eta,\alpha_s(\mu_r),Q_{\Omega})\bigg),
\end{eqnarray}
where we denote the 4-momentum transfer squared in the hard scattering as~$\hat{t}$.
We define the rapidity separation and difference of the two hard jets by
\begin{eqnarray}
\Delta\eta&=&|\eta_1-\eta_2|, \nonumber \\
\eta_{JJ}&=&\frac{1}{2}(\eta_1+\eta_2).
\end{eqnarray}
At this point we can appeal to the collinear factorisation theorems of QCD and, by working in the $\gamma p$ frame, 
write down factorised forms for the direct and
resolved cross sections. The factorised direct cross section is 
\begin{eqnarray}
\frac{\mathrm{d}\sigma_{\gamma p}^{\mathrm{dir}}}{\mathrm{d}\hat{\eta}}
(s_{\gamma p},\hat{t},\Delta\eta,\alpha_s(\mu_r),Q_{\Omega})=&&
\sum_{f_p,f_1,f_2} 
\int_{\mathrm{R_d}} \mathrm{d}x_p \, \phi_{f_p/p}(x_p,\mu_f) \nonumber \\
&&\times \frac{\mathrm{d}\hat{\sigma}^{(\gamma f)}}{\mathrm{d}\hat{\eta}}
(\hat{s},\hat{t},\Delta\eta,\alpha_s(\mu_r),Q_{\Omega},\mu_f),
\end{eqnarray}
and the factorised resolved cross section is
\begin{eqnarray}
\frac{\mathrm{d}\sigma_{\gamma p}^{\mathrm{res}}}{\mathrm{d}\hat{\eta}}
(s_{\gamma p},\hat{t},\Delta\eta,\alpha_s(\mu_r),Q_{\Omega})=&&
\sum_{f_{\gamma},f_p,f_1,f_2} \int_{\mathrm{R_r}} \mathrm{d}x_{\gamma}\,\mathrm{d}x_p 
\, \phi_{f_{\gamma}/\gamma}(x_{\gamma},\mu_f) \phi_{f_p/p}(x_p,\mu_f) \nonumber \\
&&\times \frac{\mathrm{d}\hat{\sigma}^{(f)}}{\mathrm{d}\hat{\eta}}
(\hat{s},\hat{t},\Delta\eta,\alpha_s(\mu_r),Q_{\Omega},\mu_f),
\end{eqnarray}
which are written in terms of the jet rapidity, $\hat{\eta}$, in the partonic centre-of-mass frame, and we write the 
factorisation scale and the renormalisation scale as $\mu_f$ and $\mu_r$ respectively. Note that
 $\hat{\eta}=\Delta\eta/2$, $\hat{s}=x_p W^2$ for the direct case and $\hat{s}=x_{\gamma}x_p W^2$ 
for the resolved case.
In these equations we denote the integration regions of the direct and resolved convolutions, which are defined by 
the experimental cuts, by ${\mathrm{R_d}}$ and~${\mathrm{R_r}}$. The parton distribution for a parton of flavour $f$ 
in the photon and the proton are denoted 
by $\phi_{f/\gamma}(x_{\gamma},\mu_f)$ and $\phi_{f/p}(x_p,\mu_f)$ respectively and finally 
$\frac{\mathrm{d}\hat{\sigma}^{(\gamma f)}}{\mathrm{d}\hat{\eta}}$ and 
$\frac{\mathrm{d}\hat{\sigma}^{(f)}}{\mathrm{d}\hat{\eta}}$ are the hard scattering functions which, at lowest order, start
from the Born cross sections. These are the functions that will contain the logarithmic enhancements of~$Q_{\Omega}/Q$, 
and hence depend on the definition of the gap~$\Omega$ and the gap energy flow~$Q_{\Omega}$. We
assume that~$Q_{\Omega}$ is sufficiently soft that we can ignore the effects of emission on the parent jet, known as 
recoil, but large enough that~$Q_{\Omega}^2 \gg \Lambda_{QCD}^2$.
The index~$f$ denotes the process~$f_{\gamma}+f_p \rightarrow f_1 + f_2$ and the index~$f\gamma$ denotes the 
process~$\gamma+f_p \rightarrow f_1 + f_2$. Since the aim of this paper is to calculate ratios of cross sections and 
compare with data, we 
will take the renormalisation scale to equal the factorisation scale and set~$\mu_f=\mu_r=p_t$, where $p_t$ is the
transverse momentum of the produced jets.

\subsection{Refactorisation}

Following \cite{Berger:2001ns,Kidonakis:1998bk} we now refactorise the $2\rightarrow 2$ hard scattering 
function into a hard matrix and a soft matrix,
\begin{eqnarray}
\frac{d\hat{\sigma}^{(f)}}{d\hat{\eta}}(\hat{s},\hat{t},\Delta\eta,\alpha_s(\mu_r),Q_{\Omega},\mu_f)=
&& \sum_{L,I} H^{(f)}_{IL}(\hat{s},\hat{t},\Delta\eta,\alpha_s(\mu_r),\mu_f,\mu) \nonumber \\
&& \times S^{(f)}_{LI}(Q_{\Omega},\alpha_s(\mu_r),\mu).
\end{eqnarray}
We introduce a factorisation scale~$\mu$, separate to the parton distribution factorisation scale~$\mu_f$, and all dynamics
at scales less than~$\mu$ are factored into~$S_{LI}$. Therefore $H_{IL}$ is~$Q_{\Omega}$ independent, and all 
the~$Q_{\Omega}$ dependence is included in~$S_{LI}$. This latter function describes the soft gluon dynamics. 
The proof of this statement follows standard factorisation arguments~\cite{Kidonakis:1998bk}. 
The indices~$I$ and~$L$ label the basis of colour tensors which describe the possible colour exchange in the hard 
scattering, over which the hard and soft matrices are expressed.
Soft, wide angle 
radiation decouples from the
dynamics of the hard scattering and can be approximated by an effective cross section and in this effective theory
the partons are treated as recoilless sources of gluonic radiation and replaced by eikonal lines, or path ordered
exponentials of the gluon field~\cite{Kidonakis:1998nf}. The soft radiation pattern of this effective eikonal theory 
then mimics the radiation
pattern of the partons participating in the hard event, or in other words the effective eikonal theory will
contain the same logarithms of the soft scale as the full theory. The hard scattering function will begin at order
$\alpha_s^2$ for the resolved process and order $\alpha\alpha_s$ for the direct process, and the soft function will 
begin at zeroth order. The lowest order soft function, denoted 
$S^{0}_{LI}$, reduces to a set of colour traces. Note that the definition of the gap, and hence the soft function, depends 
on the jet separation $\Delta\eta$ but we have suppressed this argument of the soft function for clarity.

The construction of the soft function, and in particular its 
renormalisation properties, have been extensively studied 
elsewhere \cite{Kidonakis:1998nf,Kidonakis:1998bk}. A non-local operator is constructed from a product of Wilson lines, which ties four lines (representing 
the four jet process) together with a colour tensor. This operator, which contains ultraviolet divergences and hence 
requires renormalisation, is used to construct a so-called eikonal cross section, which serves as an effective theory
for the soft emission. By summing over intermediate states the eikonal cross section is free of potential collinear singularities.
It is the ultraviolet renormalisation of the eikonal operator that allows colour mixing and the resummation of soft 
interjet logarithms.

\subsection{Factorisation leads to resummation of soft logarithms}

The partonic cross section, which has been factorised into a hard and a soft function, should not depend
on the choice of the factorisation scale $\mu$. This leads to the soft function obeying
\begin{equation}
\left( \mu \frac{\partial}{\partial\mu} + \beta(g_s) \frac{\partial}{\partial g_s} \right)
\uuline{S}=
-\uuline{\Gamma_s}^{\dagger}(\hat{\eta},\Omega) \uuline{S}-\uuline{S}\uuline{\Gamma_s}(\hat{\eta},\Omega).
\label{rge}
\end{equation}
It is important to point out that we have deliberately ignored the complications of terms in this equation arising from
radiation into~$\bar{\Omega}$~\cite{Berger:2001ns}, and only include radiation emitted by the soft function directly 
into~$\Omega$. 
The implication of ignoring these non-global terms is discussed in section~\ref{secngl}, where we also describe how
to include their effect in a different way. Therefore we have never included the, technically correct,~$\bar{\Omega}$ 
argument of the soft function. The matrices~$\Gamma_s(\hat{\eta},\Omega)$ are process-dependent 
soft anomalous dimension matrices that depend on the details of the gap definition and the hard scattering. 
This equation is solved by transforming to a basis in which these matrices are
diagonal and hence we require a knowledge of the eigenvectors and eigenvalues of the soft anomalous dimension matrices.
We obtain the entries for~$\Gamma_s(\hat{\eta},\Omega)$ from the coefficients of the ultraviolet poles in the matrix of 
counterterms which
renormalise the soft function; we can write this quantity as a sum over terms from different eikonal lines each with the
form of a colour factor
multiplied by a scaleless integral:
\begin{equation}
(Z_S)_{LI}=\sum_{i,j}(Z_S^{(ij)})_{LI}=\sum_{i,j}\mathcal{C}_{LI}^{(ij)} \omega^{(ij)}.
\end{equation}
The eikonal momentum integrals are process independent, and only depend on $i$ and $j$, the eikonal lines that are
connected by the virtual gluon. The colour factor is found
from consideration of the colour flow for a given process and the basis over which the colour flow is to be decomposed.
The result is a basis- and process-dependent set of colour mixing matrices, which we have listed in 
appendix \ref{appdecomp}, 
together with our choice of bases in appendix \ref{appbases}.
The colour mixing matrices have been obtained in \cite{Berger:2001ns,Kidonakis:2000gi,Berger:2003zh} 
for all relevant subprocesses, and involves using SU(3) 
colour identities like
\begin{equation}
t^a_{ij}t^a_{kl}=\frac12\delta_{il}\delta_{kj}-\frac{1}{2N_c}\delta_{ij}\delta_{kl}
\end{equation}
for quark processes and
\begin{eqnarray}
d_{abc}&=&2\left[ \mathrm{Tr}\left(t^a t^c t^b\right)+\mathrm{Tr}\left(t^a t^b t^c\right)\right], \\
f_{abc}&=&-2i\left[ \mathrm{Tr}\left(t^a t^b t^c\right)-\mathrm{Tr}\left(t^a t^c t^b\right)\right],
\end{eqnarray}
for gluon processes, to decompose one-loop graphs over a colour basis. We use the fact that the colour
flow for a real graph is the same as the corresponding virtual graph, valid for primary emission.

Therefore we need to calculate the ultraviolet divergent contribution to the momentum function $\omega^{(ij)}$ from
all contributing eikonal graphs. Working in the Feynman gauge there are two possible sources of divergence. The first 
 is one loop eikonal graphs with a virtual gluon connecting eikonal lines $i$ and $j$. From the eikonal Feynman rules
listed in the appendix, these graphs will give a real and an imaginary contribution to~$\Gamma_s$. Note that as
we are working in the Feynman gauge the self energy diagrams~($\omega^{(ii)}$) give no contribution. The second source
of ultraviolet divergences are the real emission diagrams, when the emitted gluon is directed out of the gap. This can
produce an ultraviolet divergence in the eikonal graph as we only measure energy flow into the gap and are fully inclusive
out of the gap. Hence the virtual graphs will only depend on the relative direction of the two eikonal lines and the
real graphs will give a gap (and hence a jet algorithm) dependence. This sum over real and virtual eikonal graphs 
ensures that the soft function remains free of collinear divergences. 
The imaginary (and geometry independent) part of all
our anomalous dimension matrices can be extracted 
from~\cite{Berger:2001ns,Oderda:1999kr,Kidonakis:1998nf,Kidonakis:2000gi}, and the 
calculation for a cone-algorithm defined 
final state has been done in \cite{Oderda:1999kr}. For the latter case, 
we have re-expressed their results in accordance with our notation in appendix \ref{appcone}.

By performing the energy integral of the virtual graphs, we can combine the result with the
corresponding real graph at the integrand level and obtain a partial cancellation. We can then write the total momentum 
part as an integral over the vetoed gap region and arrive at
\begin{eqnarray}
\omega^{(ij)}&=&I_{r}^{(ij)}+I_{v}^{(ij)}\nonumber \\
&=& -g_s^2 \Delta_i \Delta_k \beta_i\cdot \beta_j \int \frac{d^{d-1}k}{2|\vec{k}|(2\pi)^{d-1}} \theta(\vec{k})
\frac{1}{(\delta_i\beta_i\cdot k)(\delta_j\beta_j\cdot k)} \nonumber \\ &+& 
\delta_i \delta_j \Delta_i \Delta_j \frac{\alpha_s}{2\pi}\frac{i\pi}{2\epsilon}(1-\delta_i \delta_j),
\end{eqnarray}
where we integrate over the gap region allowed by the kt algorithm. The notation~$\delta_i=+1(-1)$ means that the
gluon momentum, denoted~$k$, flows in the same(opposite) direction as the momentum of eikonal line~$i$, and $\Delta_i=+1$ 
if the eikonal line is a quark, or it is a gluon and the soft gluon is above it in the Feynman diagram, or~$-1$ 
if the eikonal line is an antiquark, or it is a gluon and the soft gluon is below it.~
$\beta_i$ denotes the 4-velocity of eikonal line~$i$, the function~$\theta(\vec{k})=1$ when
the vector~$\vec{k}$ is directed into the gap and in 
this paper we will use the dimensional regularisation convention~$d=4-2\epsilon$.
The finite remainder is a result of the energy veto into the gap
spoiling the real/virtual cancellation. Once we have obtained the momentum integrals for the kt defined
final state we can construct the anomalous dimension matrices using the colour mixing matrices in the appendix.
Consideration of the eigenvalues and eigenvectors of these matrices, together with the process-dependent hard and 
soft matrices (the full set of hard and soft matrices is shown in appendix~\ref{apphardsoft}) 
allows the resummed cross section to be written down,
\begin{equation}
\frac{d\hat{\sigma}^{(f)}}{d\hat{\eta}}=
\sum_{L,I} \bar{H}^{0,(f)}_{IL}\bar{S}^{0,(f)}_{LI} 
\exp\left\{ \frac{1}{\beta_0}(\hat{\lambda}_L^*(\hat{\eta},\Omega)+\hat{\lambda}_I(\hat{\eta},\Omega))
\int_{p_t}^{Q_{\Omega}} 
\frac{d\mu}{\mu}\beta_0\alpha_s(\mu)\right\},
\label{eqresum}
\end{equation}
which follows from the diagonalisation of the soft RG equation \ref{rge}. We denote matrices in the diagonal basis by barred
matrices, the eigenvalues of the anomalous dimension matrices by $\lambda_i=\alpha_s\hat{\lambda}_i$ 
and we write the lowest-order piece of the QCD beta function as $\beta_0=(11N_c-2n_f)/(6\pi)$. We will observe that, in 
agreement with \cite{Oderda:1999kr}, $\mathrm{Re}(\lambda)>0$ for all physical channels 
and hence the resummed cross sections are suppressed relative to the 
fully inclusive cross section. 

\subsection{Non-global effects}

\label{secngl}

As we have discussed in the last section, we have deliberately ignored terms arising from secondary radiation into
$\Omega$, or non-global logarithms (NGLs) \cite{Appleby:2003ai,Appleby:2002ke,Dasgupta:2002bw,Dasgupta:2001sh}. Such
terms arise from radiation at some intermediate scale,~$M$, being emitted outside of~$\Omega$, i.e.~into $\bar{\Omega}$,
and then subsequently radiating into $\Omega$. In energy flow observables such effects give rise to leading
logarithms. Inclusion of NGLs in the formalism of the last section would result in an explicit $M$ dependence of the
soft function and a sensitivity to more complicated, $2\rightarrow n$, colour flows for all $n>2$. For further details
see \cite{Berger:2001ns}.  NG effects have been studied for a two-jet system by Dasgupta and Salam
\cite{Dasgupta:2002bw,Dasgupta:2001sh} and by the current authors with the complication of clustering
\cite{Appleby:2002ke}, and in the context of energy flow/event shape correlations by Dokshitzer and Marchesini
\cite{Dokshitzer:2003uw} and Berger, K\'ucs and Sterman \cite{Berger:2003iw}.

The effect of NGLs for four-jet kinematics has not been explicitly calculated to date.  The best that has been managed
is a two-jet calculation in the large-$N_c$ limit.  The NG contributions to the gap cross section factorize into an
overall suppression factor $S^{NG}$, making it smaller than would be predicted by the resummation of primary logarithms
alone.  In the absence of a complete calculation, we include the NGLs approximately, by using our all-order results in
the large-$N_c$ limit for $S^{NG}$ in a two-jet system \cite{Appleby:2002ke}.  Since four-jet configurations are
dominated, in the large-$N_c$ limit and for large $\Delta\eta$, by colour flows in which two colour dipoles stretch
across the gap region, we approximate the four-jet NG suppression factor by the square of the two-jet one.

We have reperformed our previous calculation for the kinematic range relevant to HERA and find that the variation of
$S^{NG}$ with $\Delta\eta$ is very weak, so we neglect it.  The variation with $Q_\Omega$ is very strong on the other
hand.  $S^{NG}$~is a function of $t$,
\begin{equation}
t=\frac{1}{2\pi \beta_0}\log\left(\frac{\alpha_s(Q_{\Omega})}{\alpha_s(Q)}\right),
\end{equation}
where $\beta_0=(11 C_A-2 n_f)/(6\pi)$, and is well-approximated by a Gaussian in $t$.  Thus if $Q_\Omega$ is too close
to $\Lambda_{QCD}$, $t$ varies rapidly with it and $S^{NG}$ varies very rapidly.

\TABLE{
\begin{tabular}{|c|c|}
 \hline
$Q_{\Omega}$ [GeV] & $S^{NG}(t)^2$ \\ \hline\hline
0.5 & 0.10$^{+0.30}_{-0.10}$ \\ \hline 
1.0 & 0.47$^{+0.16}_{-0.22}$ \\ \hline
1.5 & 0.65$^{+0.10}_{-0.13}$ \\ \hline
2.0 & 0.74$^{+0.07}_{-0.08}$ \\ \hline
\end{tabular}
\caption{The non-global emission suppression factors for the 4-jet system, obtained from an all-orders
calculation for $Q=6$~GeV.}
\label{nglsupp}
}
It is impossible to quantify the uncertainties in this approximation, without a more detailed understanding of the
underlying physics.  To get an idea however, we estimate the possible size of higher order corrections, by varying the
hard scale at which $\alpha_s$ is evaluated.  To leading logarithmic accuracy, this is equivalent to varying the value
of $\alpha_s(Q)$ by an amount proportional to its value.  We therefore evaluate $t$, and hence $S^{NG}(t)^2$, using our
central value of $\alpha_s(M_z)=0.116$, which implies $\alpha_s(Q=\mbox{6~GeV})=0.196$, and with raised and lowered
values $\alpha_s^{\mathrm{up}}(\mbox{6~GeV})=0.234$ and $\alpha_s^{\mathrm{down}}(\mbox{6~GeV})=0.158$.  For
$Q_\Omega=1.0$~GeV, for example, these values result in $t=0.097^{+0.056}_{-0.032}$ and hence
$S^{NG}(t)^2=0.47^{+0.16}_{-0.22}$.  We show the results for all relevant values of $Q_\Omega$ in table \ref{nglsupp}.
Note that $Q_\Omega=0.5$~GeV is so low that the range of uncertainty in $t$ extends beyond $\Lambda_{QCD}$ and hence the
estimate of $S^{NG}$ extends to zero.  We have not shown any results for the 1995 cone-based ZEUS energy flow analysis
\cite{Derrick:1995pb} because the low value of $Q_{\Omega}=0.3$~GeV means that the central value of the NG suppression
is already zero, indicating a breakdown of our perturbative approach.

The uncertainty on the secondary emission probability estimated in this way should be added to that on the primary
emission probability, described in section \ref{secresults}.  However, we will find that the secondary uncertainty
generally dominates the two.  This is therefore clearly an area that needs more work if more precise quantitative
predictions are to be made.

\section{Soft gluon dynamics for a \boldmath{kt} defined final state}

\label{seckt}

We now evaluate the momentum integral, $\omega^{(ij)}$, over the gap region $\Omega$. 
The region of integration is determined by the experimental geometry, in which the final state is defined
by the kt algorithm, and we shall work with the quantity
\begin{equation}
\Omega_{kt}^{(ij)}=\int_{kt} d\eta \int_{kt} \frac{d\phi}{2\pi}
\frac{\beta_i\cdot \beta_j}{(\beta_i\cdot \bar{k})(\beta_j\cdot \bar{k})},
\end{equation}
where we define $\bar{k}=k/k_t$. Therefore
\begin{equation}
\omega^{(ij)}=-\frac{\alpha_s}{2\pi} \Delta_i \Delta_j \delta_i \delta_j \frac{1}{2\epsilon}\Omega_{kt}^{(ij)}+I.P..
\end{equation}  
We denote the geometry independent imaginary part by $I.P.$, and we define the finite piece $\Gamma^{(ij)}$ by
\begin{equation}
\omega^{(ij)}=-\mathcal{S}_{ij}\frac{\Gamma^{(ij)}}{2\epsilon}.
\end{equation}
We have extracted the sign function from $\Gamma^{(ij)}$,
\begin{equation}
\mathcal{S}_{ij}=\Delta_i \Delta_j \delta_i \delta_j,
\label{eqsign}
\end{equation}
so that 
\begin{equation}
\Gamma^{(ij)}=\frac{\alpha_s}{2\pi}\Omega^{(ij)}_{kt}+I.P..
\end{equation}
In this work we denote the rapidity separation of the jets by 
$\Delta\eta$ and the width of an azimuthally symmetric rapidity gap by $\Delta y$ ($<\Delta\eta$). Therefore the available 
phase space for soft gluon emission for a kt defined final state is given by
\begin{equation}
\Omega_{kt}^{(ij)}=\lim_{\Delta y \rightarrow \Delta\eta} \left(
\Omega^{(ij)}_f(\Delta y,\Delta\eta) - \Omega^{(ij)}_1(\Delta y,\Delta\eta,R) - 
\Omega^{(ij)}_2(\Delta y,\Delta\eta,R)\right),
\end{equation}
where the first term arises from an azimuthally symmetric gap of width $\Delta y$, and we subtract the region around 
each jet which is vetoed by the kt algorithm. The regions of this equation are shown in figure \ref{ktphasespace}. 
\FIGURE{
\begin{minipage}{0.9\textwidth}
\begin{center}
\includegraphics*[width=7cm,height=6cm]{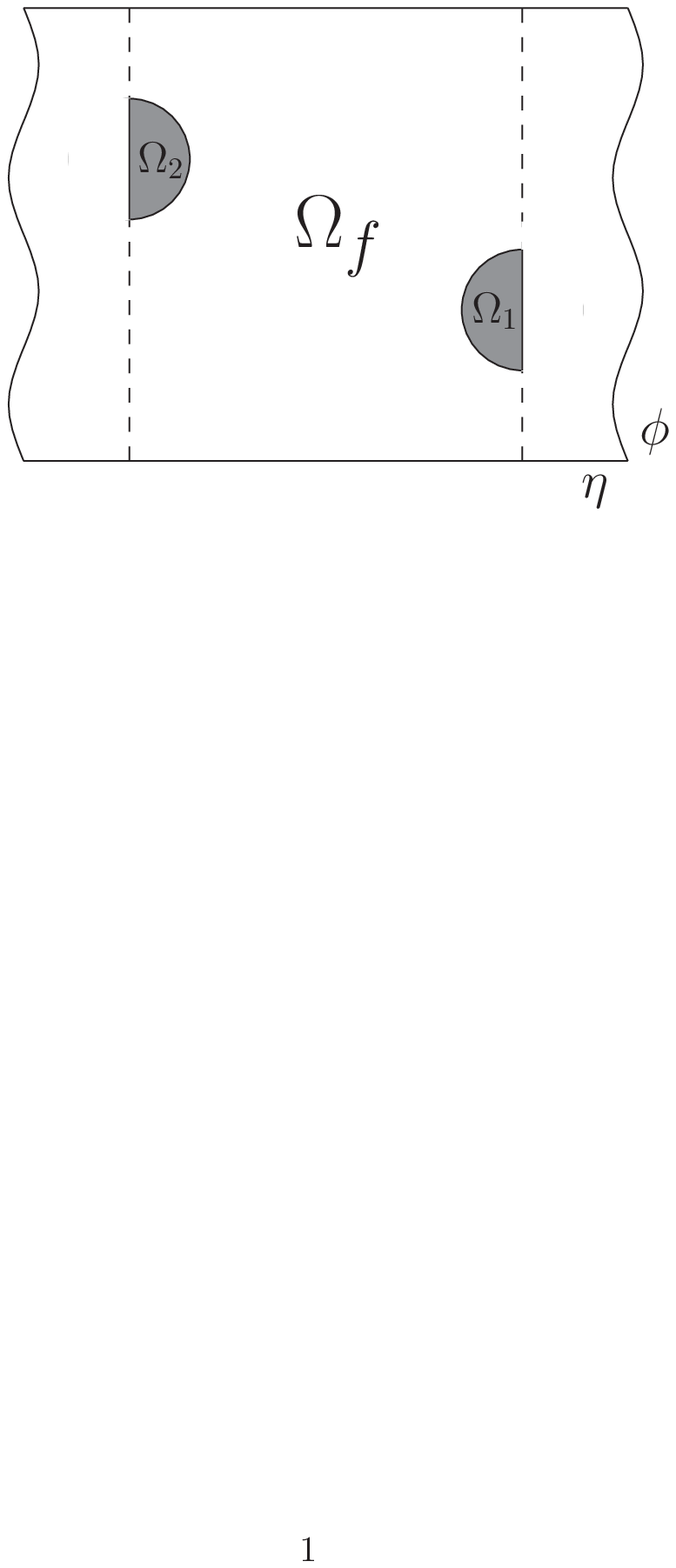}
\end{center}
\end{minipage}
\caption{The phase space regions for a kt defined final state. The shading denotes the regions 
vetoed by the algorithm, which are subtracted from the $\Omega_f$ piece. Note that we have dropped 
the ${(ij)}$ superscripts in this figure.}
\label{ktphasespace}}
In these regions any soft radiation is clustered into the jet, and cannot 
form part of $\Omega$. In the first term we take~$\Delta y$ approaching~$\Delta\eta$, 
and hence it contains 
a collinear divergence when the emitted gluon is collinear to one of the jets. The two subtracted pieces then
 remove the regions of phase space defined by
\begin{equation}
(\eta_k-\eta_i)^2+(\phi_k-\phi_i)^2<R^2,
\end{equation}
where the index $i$ labels final state jets and $k$ labels the emitted gluon. 
The collinear divergences in the subtracted pieces exactly match the 
divergences in the first piece and hence the function $\Omega^{(ij)}_{kt}(\Delta\eta)$ is
collinear safe. Explicit definitions of the $\Omega$ functions are
\begin{eqnarray}
\Omega^{(ij)}_f&=&\int_{-\Delta y/2}^{+\Delta_y/2}d\eta \int_0^{2\pi} \frac{d\phi}{2\pi} 
\frac{\beta_i\cdot \beta_j}{(\beta_i\cdot \bar{k})(\beta_j\cdot \bar{k})}, \nonumber \\
\Omega^{(ij)}_1&=&\int_{\Delta\eta/2-R}^{+\Delta y/2}d\eta 
\int_{-\phi_{\mathrm{lim}}}^{+\phi_{\mathrm{lim}}} d\phi 
\frac{\beta_i\cdot \beta_j}{(\beta_i\cdot \bar{k})(\beta_j\cdot \bar{k})},
\end{eqnarray}
where we write $\phi_{\mathrm{lim}}=\sqrt{R^2-(\eta-\Delta\eta/2)^2}$ and obtain $\Omega^{(ij)}_2$ by
the symmetry $\Omega^{(ij)}_2 = \Omega^{(\bar\imath\bar\jmath)}_1$, where the mapping $i\to\bar\imath$
is given by $\{a,b,1,2\}\to\{b,a,2,1\}$. If we define the following combinations of momentum integrals,
\begin{eqnarray}
\alpha&=&\mathcal{S}_{ab}\Gamma^{(ab)}+\mathcal{S}_{12}\Gamma^{(12)}, \nonumber \\
\beta&=&\mathcal{S}_{a1}\Gamma^{(a1)}+\mathcal{S}_{b2}\Gamma^{(b2)}, \nonumber \\
\gamma&=&\mathcal{S}_{b1}\Gamma^{(b1)}+\mathcal{S}_{a2}\Gamma^{(a2)},
\end{eqnarray}
where we have combined classes of diagram with the same colour structure, we obtain the following closed form 
for the positive gap contributions, in the limit $\Delta y \rightarrow \Delta\eta$,
\begin{eqnarray}
\alpha &=&
\frac{\alpha_s}{\pi}\Bigl(\phantom{-}2\Delta\eta+
\log\bigl(1-e^{-2\Delta\eta}\bigr)+\log\frac1{\Delta\eta-\Delta y}
-2i\pi\Bigr), \phantom{(10)} \label{eqalpha} \\
\beta &=&
\frac{\alpha_s}{\pi}\Bigl(\phantom{-2\Delta\eta+{}}
\log\bigl(1-e^{-2\Delta\eta}\bigr)+\log\frac1{\Delta\eta-\Delta y}
\Bigr), \\
\gamma &=&
\frac{\alpha_s}{\pi}\Bigl(-2\Delta\eta-
\log\bigl(1-e^{-2\Delta\eta}\bigr)-\log\frac1{\Delta\eta-\Delta y}
\Bigr).
\label{eqgamma}
\end{eqnarray}
The subtraction pieces are straightforward to 
express as power series in $R$ and $e^{-\Delta\eta}$ and we shall illustrate the calculation of the momentum 
integrals with an example.

\subsection{Calculation of \boldmath{$\Omega_{kt}^{(a1)}$}} 

We can write the matrix element in terms of the rapidity of the emitted gluon and obtain the 
following matrix element
\begin{equation}
\frac{\beta_i\cdot \beta_j}{(\beta_i\cdot \bar{k})(\beta_j\cdot \bar{k})}=
\frac{e^{-\Delta\eta/2}}{e^{-\eta}(\cosh(\Delta\eta/2-\eta)-\cos\phi)}.
\end{equation}
The integrations for the function $\Omega_f^{(a1)}$ are straightforward, and we obtain
\begin{equation}
\Omega_f^{(a1)}=-\Delta y + \log\left(\frac{\sinh(\Delta\eta/2+\Delta y/2)}{\sinh(\Delta\eta/2-\Delta y/2)}\right).
\end{equation}
The expression for $\Omega_1^{(a1)}$ is
\begin{eqnarray}
\Omega^{(a1)}_1&=&\int_{\Delta\eta/2-R}^{+\Delta y/2}d\eta 
\int_{-\phi_{\mathrm{lim}}}^{+\phi_{\mathrm{lim}}} \frac{d\phi}{2\pi} 
\frac{e^{-\Delta\eta/2}}{e^{-\eta}(\cosh(\Delta\eta/2-\eta)-\cos\phi)}, \nonumber \\
&=&\int_{\Delta\eta/2-R}^{+\Delta y/2}d\eta \, f(\eta,\Delta\eta,R), \nonumber \\
&=&\int_{\Delta\eta/2-\Delta y/2}^{R}d\eta' \, f(\eta',\Delta\eta,R),
\end{eqnarray}
where $\phi_{\mathrm{lim}}$ is defined in the previous section, we have performed the azimuthal 
integration in the second step and changed variable to $\eta'=\Delta\eta/2-\eta$ in the third step. 
The function $f$ can be easily obtained, but it is rather lengthy so we do not reproduce it here. 
We now note that this expression for $\Omega^{(ij)}_1$ only involves jet 1 and hence, by Lorentz
invariance, cannot depend on the other jet and so may not be a function of the jet separation 
$\Delta\eta$. Therefore we write
\begin{equation}
\Omega^{(a1)}_1=\int_{\Delta\eta/2-\Delta y/2}^{R}d\eta' \, f(\eta',R).
\end{equation}
This function $f(\eta',R)$ has a divergence as $\eta'\rightarrow 0$, so we add and subtract this 
divergence to obtain
\begin{equation}
\Omega^{(a1)}_1=\int_{\Delta\eta/2-\Delta y/2}^R d\eta' \left(f(\eta',R)-\frac{1}{\eta'}\right) 
+ \int_{\Delta\eta/2-\Delta y/2}^R \frac{d\eta'}{\eta'}.
\end{equation}
We can rewrite the lower limit of the first, divergence free, integral as 0, and the collinear
divergence is now contained in the second term. Therefore we have used $\Delta y$ as a cut-off for
the divergence, and we can write
\begin{equation}
\Omega^{(a1)}_1=\bar{\Omega}^{ij}_1+\log{2R}-\log(\Delta\eta-\Delta y).
\end{equation}
We will always denote the divergence free angular integration, which always results
from such a subtraction, as a barred quantity. We can now rescale the $\bar{\Omega}^{(a1)}_1$ integral,
using $\bar{\eta}=\eta'/R$, to obtain
\begin{equation}
\bar{\Omega}^{(a1)}_1=\int_0^1 d\bar{\eta} \left(R\cdot g(\bar{\eta},R)-\frac{1}{\bar{\eta}}\right).
\end{equation}
This quantity, which is only a function of $R$, can now be expressed as a power series in $R$ and the
integrals done on a term-by-term basis. Doing this we obtain the rapidly converging series,
\begin{equation}
\bar{\Omega}^{(a1)}_1=-\log(2)-\frac{2R}{\pi}+\frac{R^2}{8}-\frac{R^3}{18\pi}+\frac{R^4}{576}
-\frac{R^5}{5400\pi}-\frac{R^7}{529200\pi}+\frac{R^8}{4147200}+\dots.
\end{equation}
To calculate $\Omega^{(a1)}_2$ we use the parity symmetry and obtain the expression,
\begin{eqnarray}
\Omega^{(a1)}_2&=&\Omega^{(b2)}_1 \nonumber \\
&=&\int_{\Delta\eta/2-R}^{+\Delta y/2}d\eta 
\int_{-\phi_{\mathrm{lim}}}^{+\phi_{\mathrm{lim}}} \frac{d\phi}{2\pi} 
\frac{e^{-\Delta\eta/2}}{e^{\eta}(\cosh(\Delta\eta/2+\eta)+\cos\phi)}.
\end{eqnarray}
We now perform similar manipulations to the case of $\Omega^{(a1)}_1$. However, as $\Omega^{(a1)}_2$
is a function of both final state jets, the resulting expression must be a function of $\Delta\eta$ and
we also note that $\Omega^{(a1)}_2$ is not divergent. We hence obtain the expression
\begin{equation}
\bar{\Omega}_2^{(a1)}=\int_0^1 d\bar{\eta}\left(
R\cdot f(\bar{\eta},\Delta\eta,R)\right),
\end{equation}
which we can expand as a power series in the variables $R$ and $z=\exp(-\Delta\eta)$, and perform
the remaining integrations term-by-term.

The pole arising in the subtraction term $\Omega_1^{(a1)}$ now cancels against an equivalent pole
in the function $\Omega_f^{(a1)}$, when we expand the latter in $\Delta y$ around the point $\Delta\eta$,
\begin{equation}
\lim_{\Delta y\rightarrow \Delta\eta}\Omega_f^{(a1)}
\sim-\Delta \eta -\log(\Delta\eta-\Delta y) + \log(2\sinh\Delta\eta).
\end{equation}
Therefore we find the final, divergence free, form of $\Omega_{kt}^{(a1)}$ as
\begin{equation}
\Omega_{kt}^{(a1)}=-\Delta\eta+\log(2\sinh\Delta\eta)-\log(2R)-\bar{\Omega}_1^{(a1)}-\bar{\Omega}_2^{(a1)}.
\end{equation}
We have presented the full set of series expansions in appendix \ref{appgamma} and these, together with 
equations \ref{eqalpha}--\ref{eqgamma}, are sufficient to compute the set of kt defined momentum integrals and hence the
corresponding anomalous dimension matrix. 
It is worth noting that, although the off-diagonal terms for the kt anomalous dimension matrices are no longer
pure imaginary as in the cone case, their real parts still vanish for large~$\Delta\eta$.
Indeed for~$\Delta\eta=2$, the real part is more than two orders
of magnitude smaller than the imaginary part. We have listed the
 closed-form momentum integrals for the cone defined final state using our notation 
in appendix~\ref{appcone}.

\section{Results}

\label{secresults}

We now have the tools we need to calculate resummed cross sections at HERA, which correctly include primary emission 
to all orders and secondary emission approximately in the large $N_c$ limit. The colour bases
used for the contributing partonic cross sections are presented in the appendix, along with 
the decomposed hard and soft matrices. We also present the complete colour mixing matrices and the correct 
sign structure for the three classes of diagram. Therefore we can use the eigenvectors and eigenvalues
of the soft anomalous dimension matrices, together with the hard and soft matrices, to calculate the primary 
resummed cross section using equation~\ref{eqresum}, for either a kt or a cone defined final state. 
The differential cross section, in~$\Delta\eta$, 
can then be computed using the cuts given in section \ref{sechera}, both for the totally inclusive cross section
 (no gap) and for the gap cross section at fixed $Q_{\Omega}$. The gap fraction is then found by dividing the latter
quantity by the former. All our results are computed using GRV photon parton densities \cite{Gluck:1991jc} and 
the MRST proton parton densities \cite{Martin:1998np}. We have included an 
estimate of the theoretical
uncertainty in the primary resummation by varying the hard scale in the evaluation of $\alpha_s$, while keeping the 
ratio of the hard and soft scales fixed.

\subsection{Totally inclusive \boldmath{ep} cross section and the gap cross section}

\FIGURE{\\
\begin{minipage}{0.48\textwidth}
\includegraphics*[width=7cm,height=6cm]{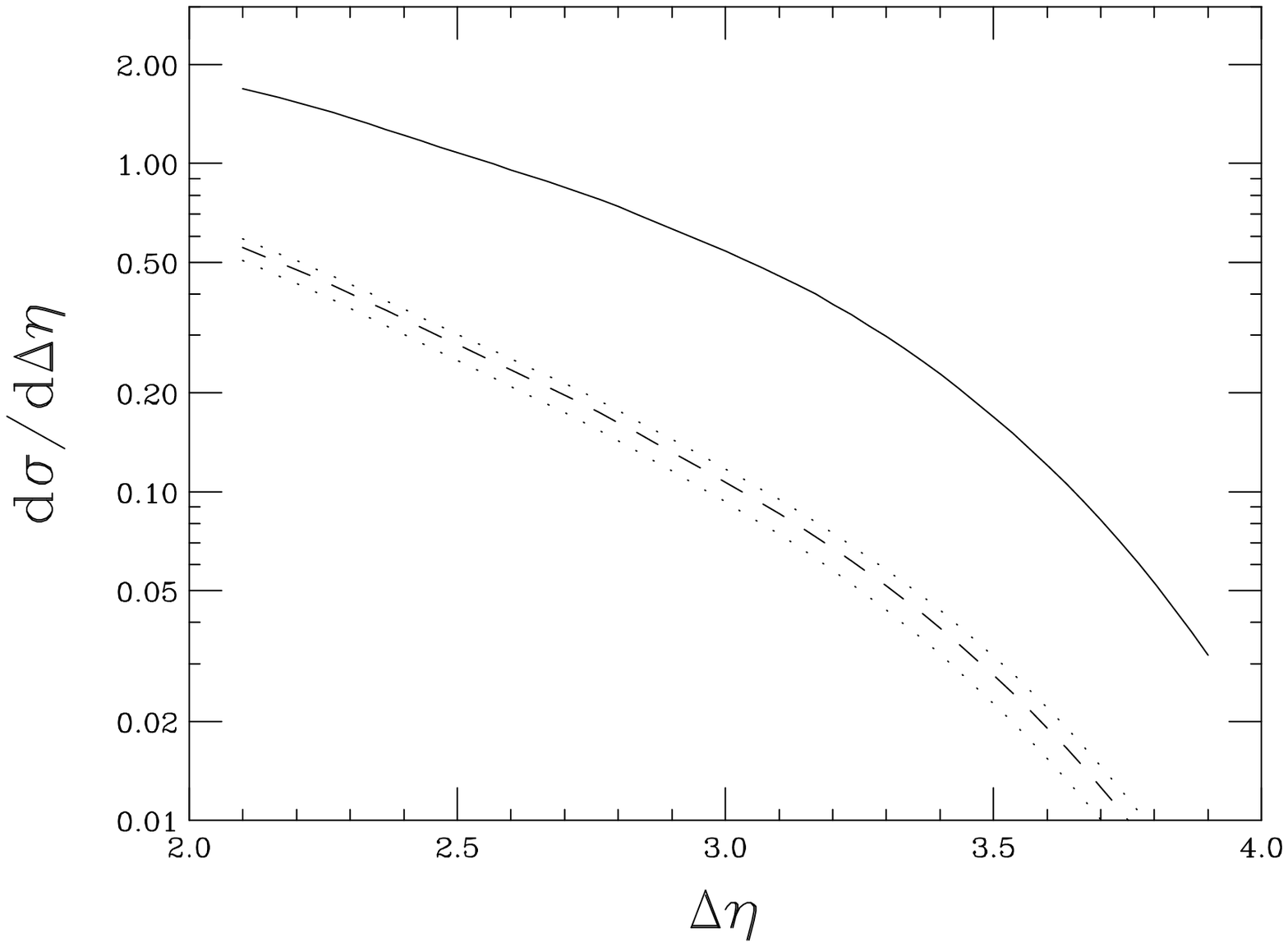}
\end{minipage}
\hfill
\begin{minipage}{0.48\textwidth}
\includegraphics*[width=7cm,height=6cm]{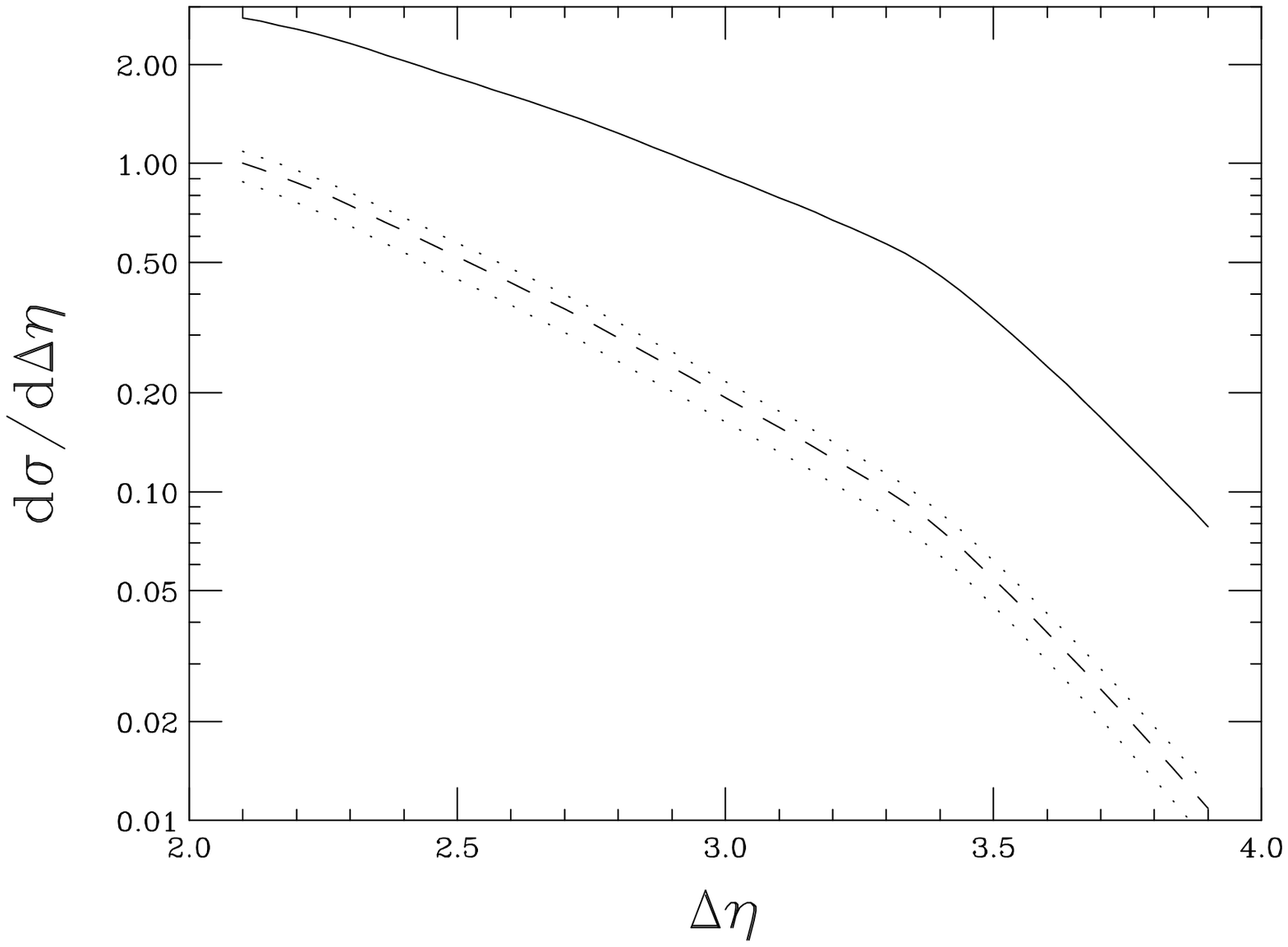}
\end{minipage}
\label{figgap}
\caption{The cross sections for the H1 data (left) and the ZEUS data (right), which was defined using the kt
algorithm with $R=1.0$. On both plots the solid line is the total 
inclusive cross section, the dashed line is the gap cross section
for $Q_{\Omega}=1$~GeV with only primary emission included, and the dotted lines indicate the range of 
theoretical uncertainty in the prediction.}}

\FIGURE{\\
\begin{minipage}{0.48\textwidth}
\includegraphics*[width=7cm,height=6cm]{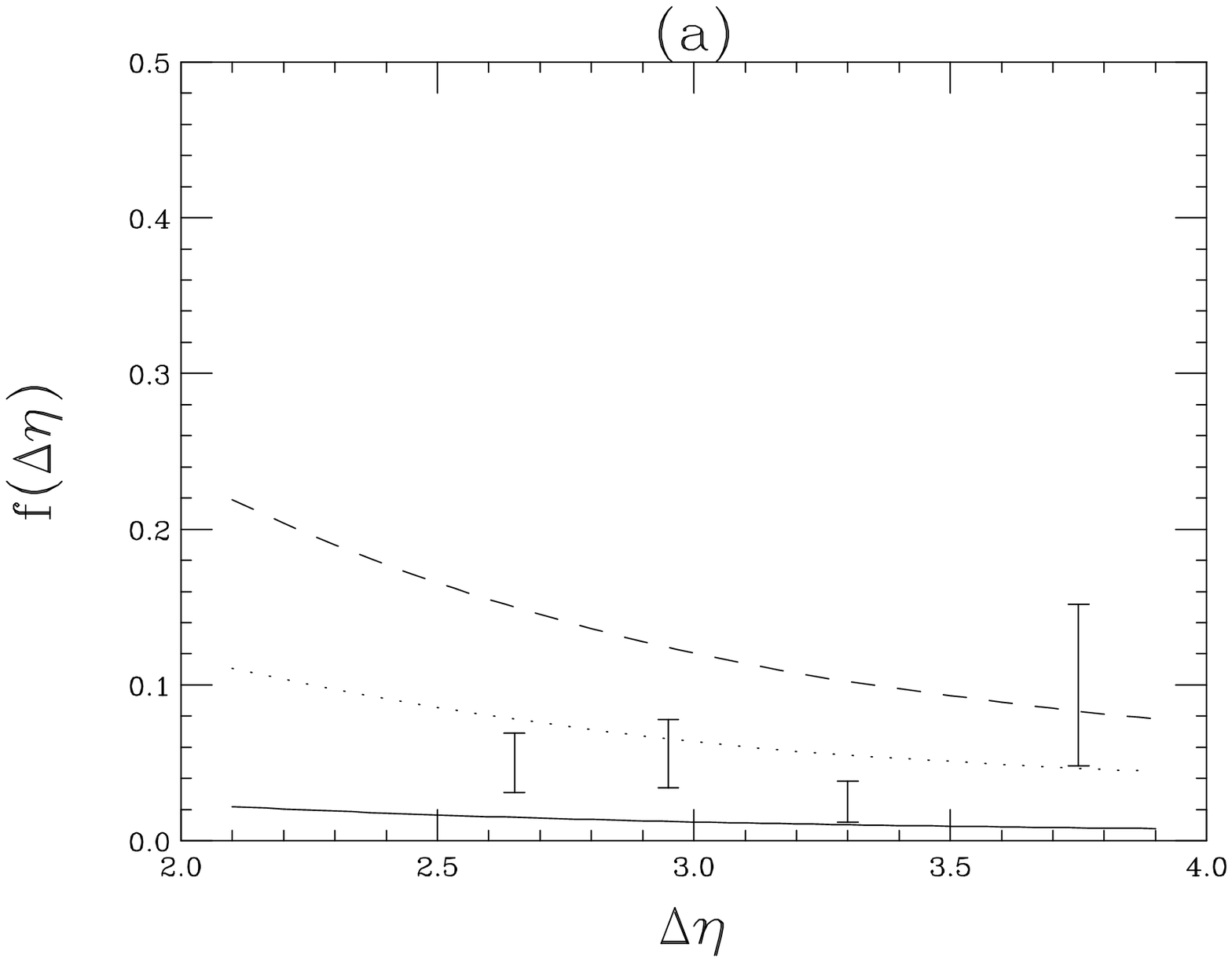}	
\end{minipage}
\hfill
\begin{minipage}{0.48\textwidth}
\includegraphics*[width=7cm,height=6cm]{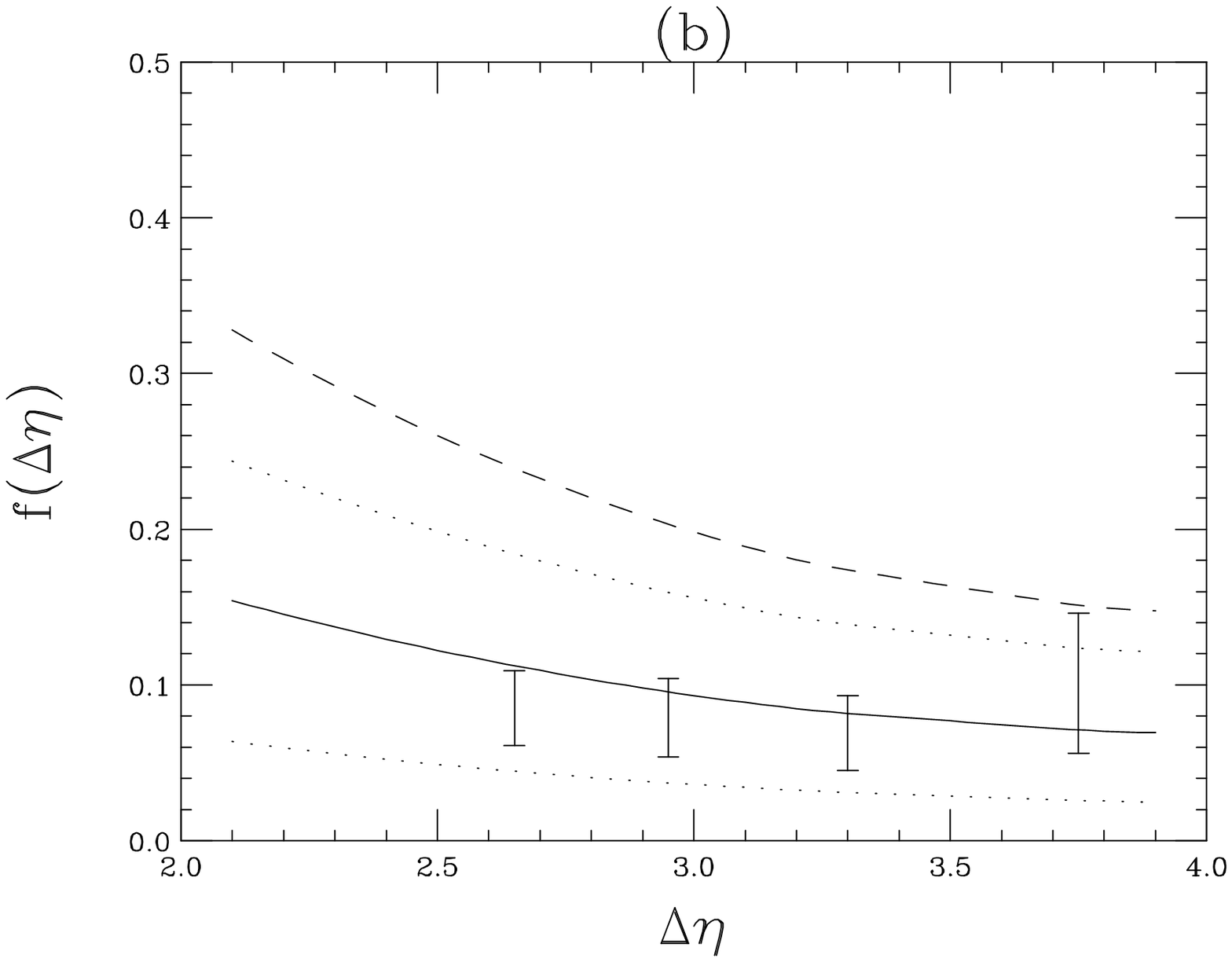}
\end{minipage}\\
\begin{minipage}{0.48\textwidth}
\includegraphics*[width=7cm,height=6cm]{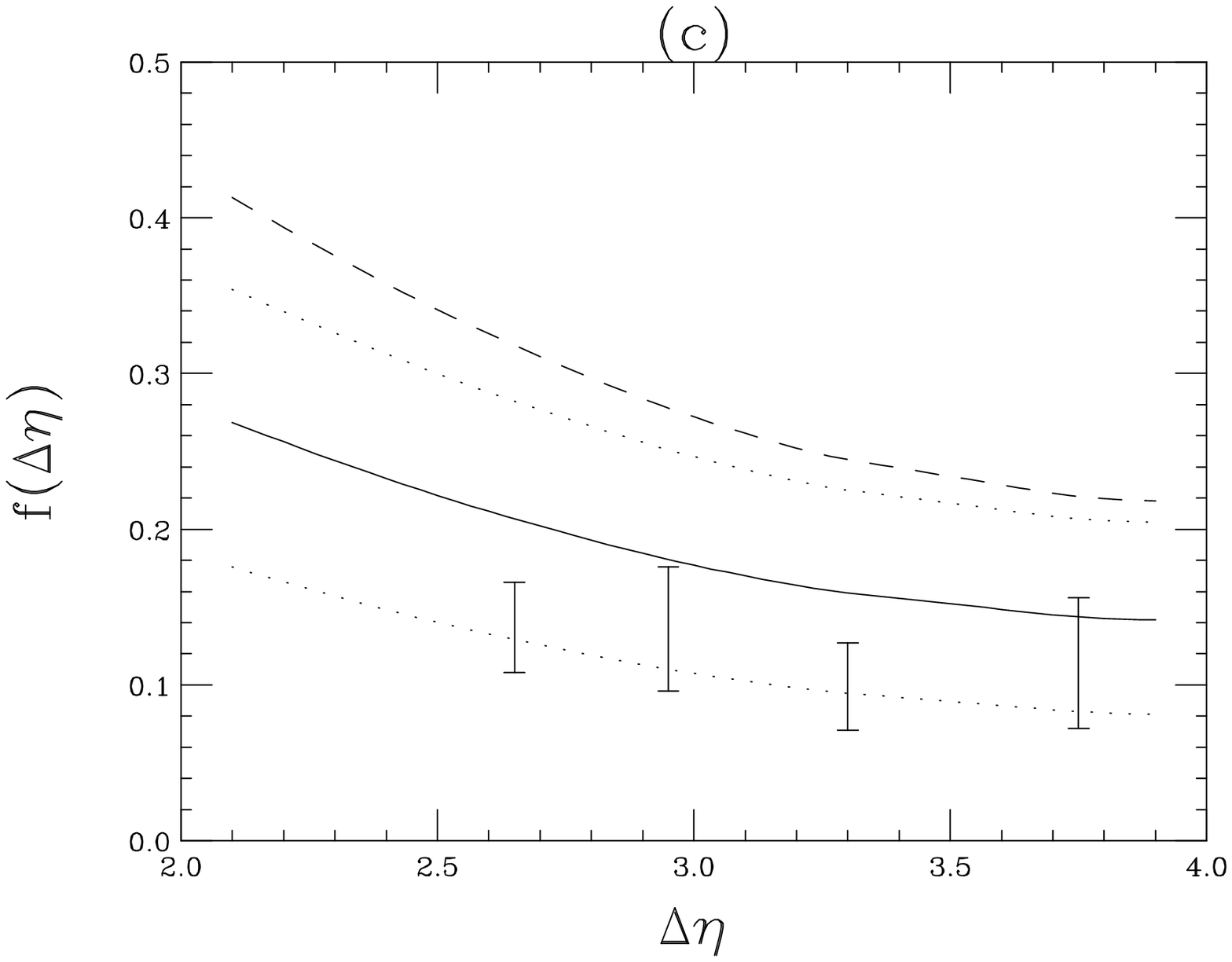}	
\end{minipage}
\hfill
\begin{minipage}{0.48\textwidth}
\includegraphics*[width=7cm,height=6cm]{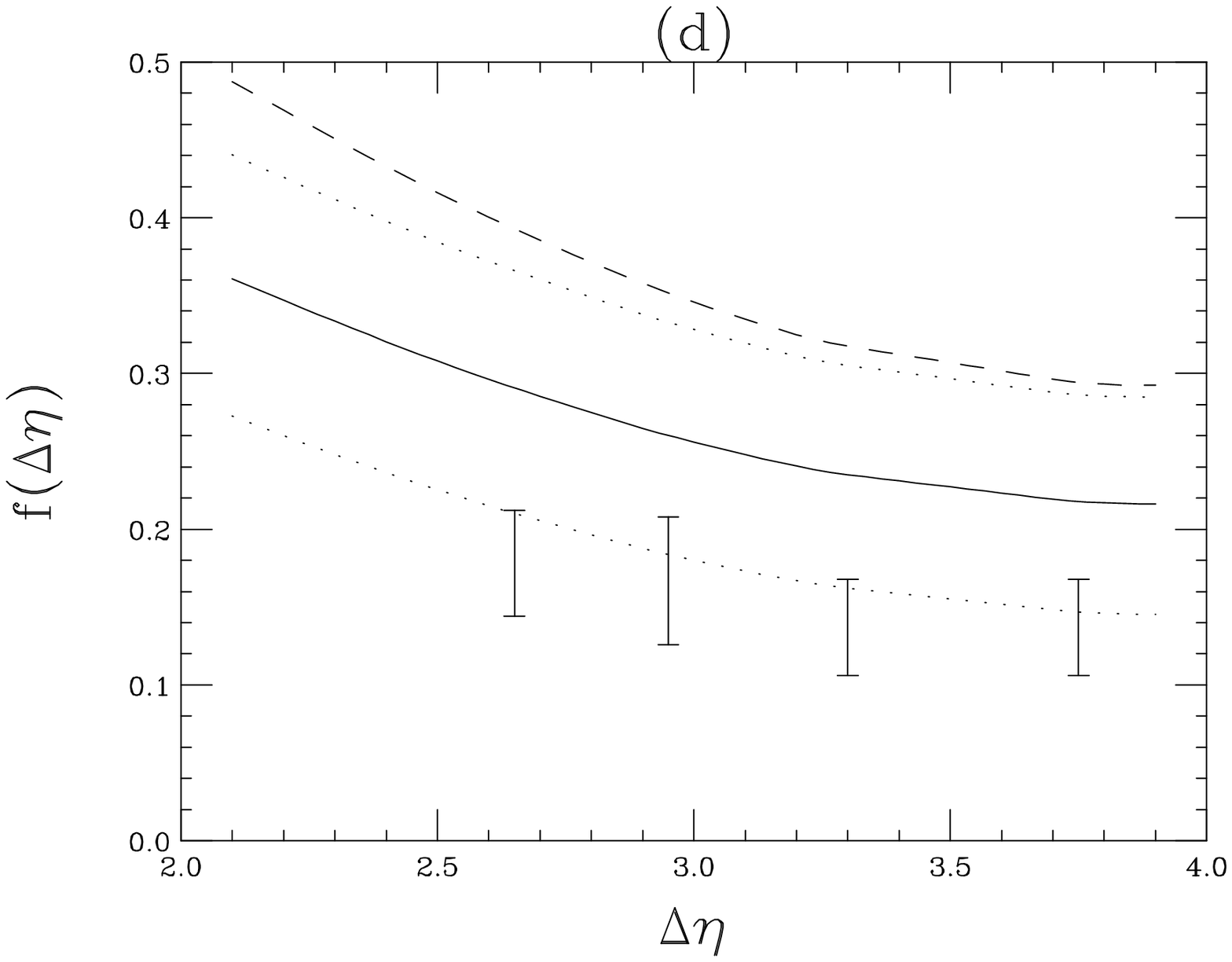}
\end{minipage}\\
\caption{The gap fractions for the H1 analysis with a kt defined final state ($R=1.0$), 
at varying $Q_{\Omega}$. $Q_{\Omega}=0.5,\,1.0,\,1.5,\,2.0$~GeV for plots (a), (b), (c) and (d) respectively.
The H1 data is also shown. The solid line includes the effects of primary emission and 
the secondary emission suppression factor. The overall theoretical
uncertainty, including the primary uncertainty and the secondary uncertainty, is shown by the
dotted lines. The dashed line indicates the gap fraction obtained by including only primary emission.}
\label{figh1frac}}

\FIGURE{\\
\begin{minipage}{0.48\textwidth}
\includegraphics*[width=7cm,height=6cm]{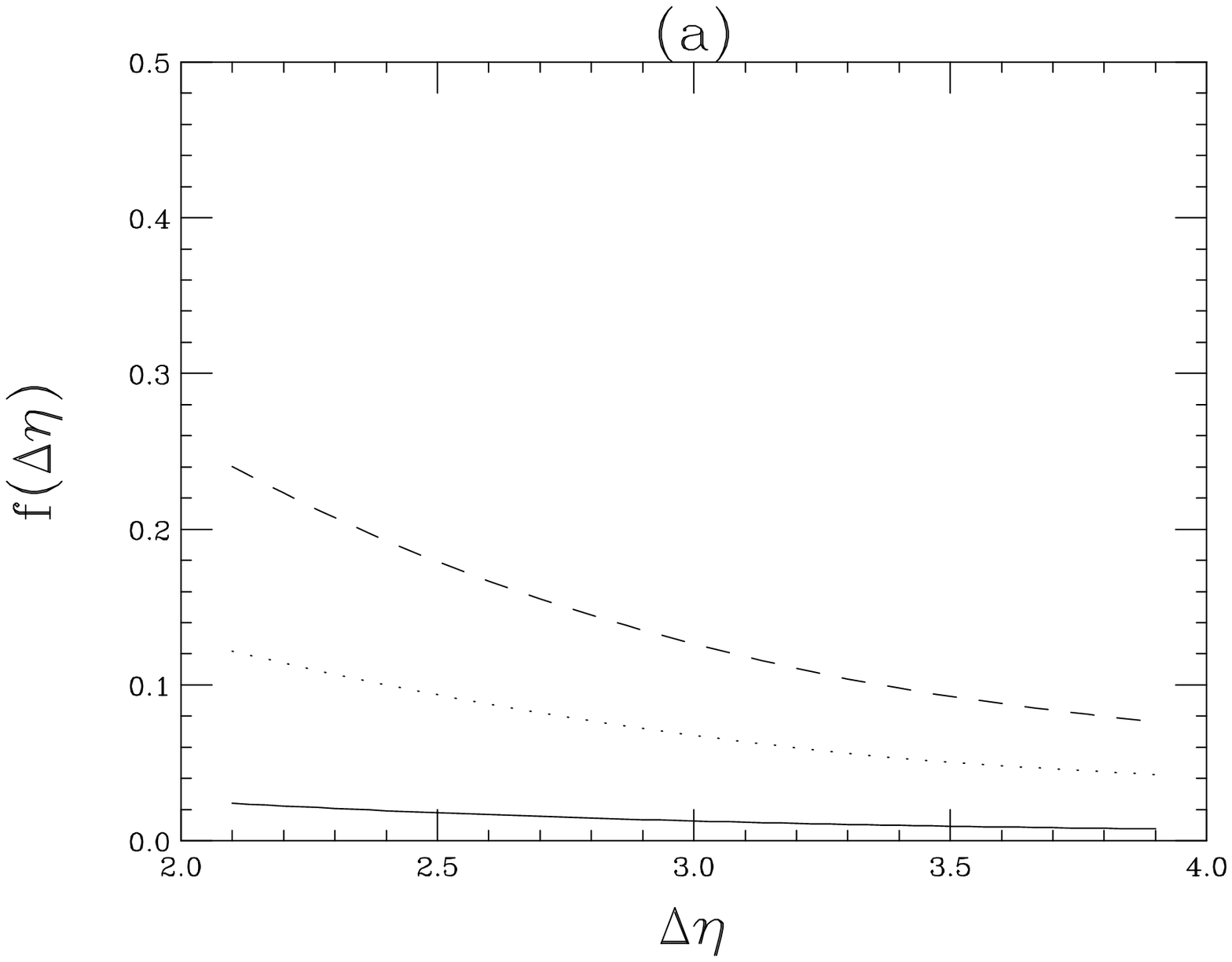}	
\end{minipage}
\hfill
\begin{minipage}{0.48\textwidth}
\includegraphics*[width=7cm,height=6cm]{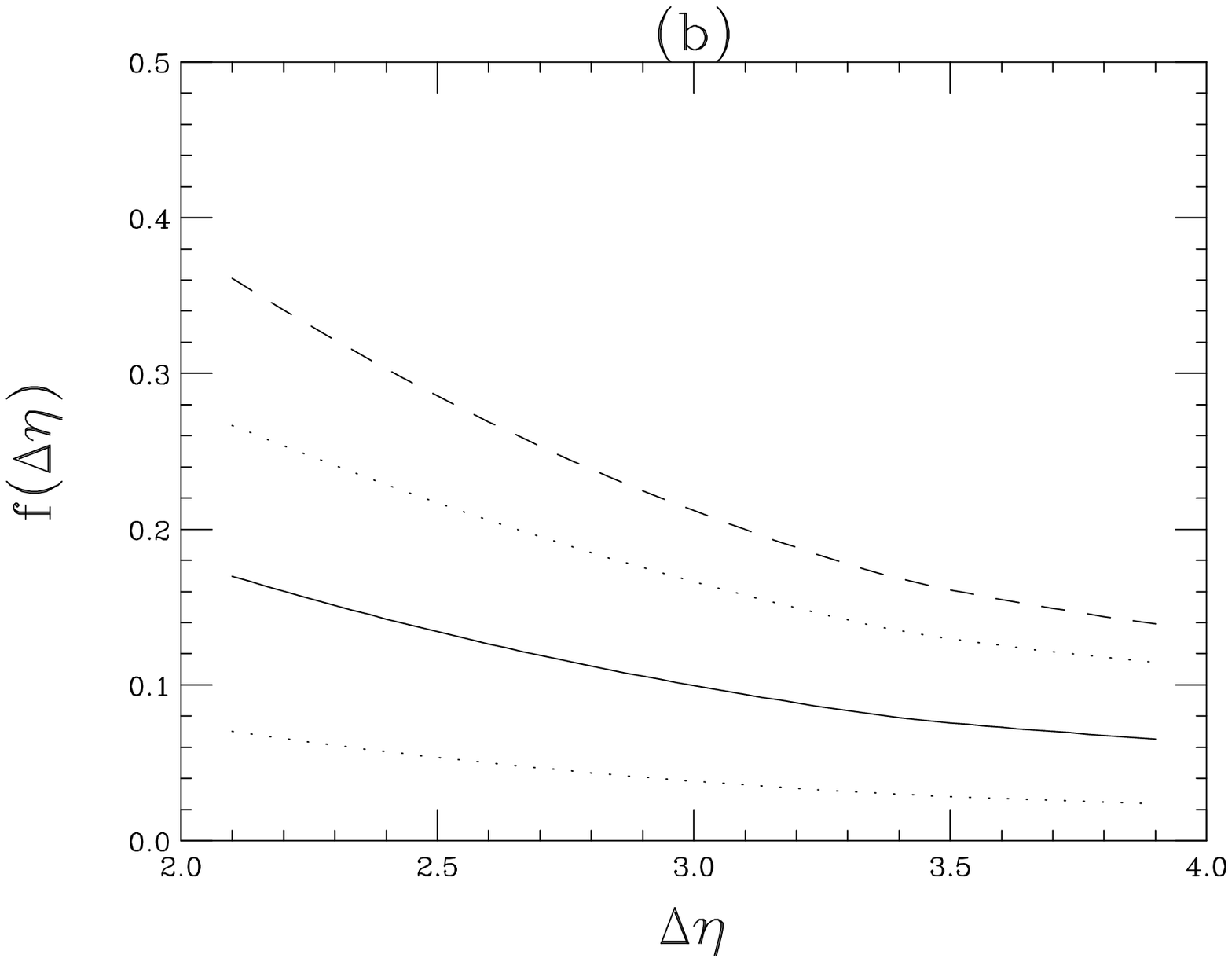}
\end{minipage}\\
\begin{minipage}{0.48\textwidth}
\includegraphics*[width=7cm,height=6cm]{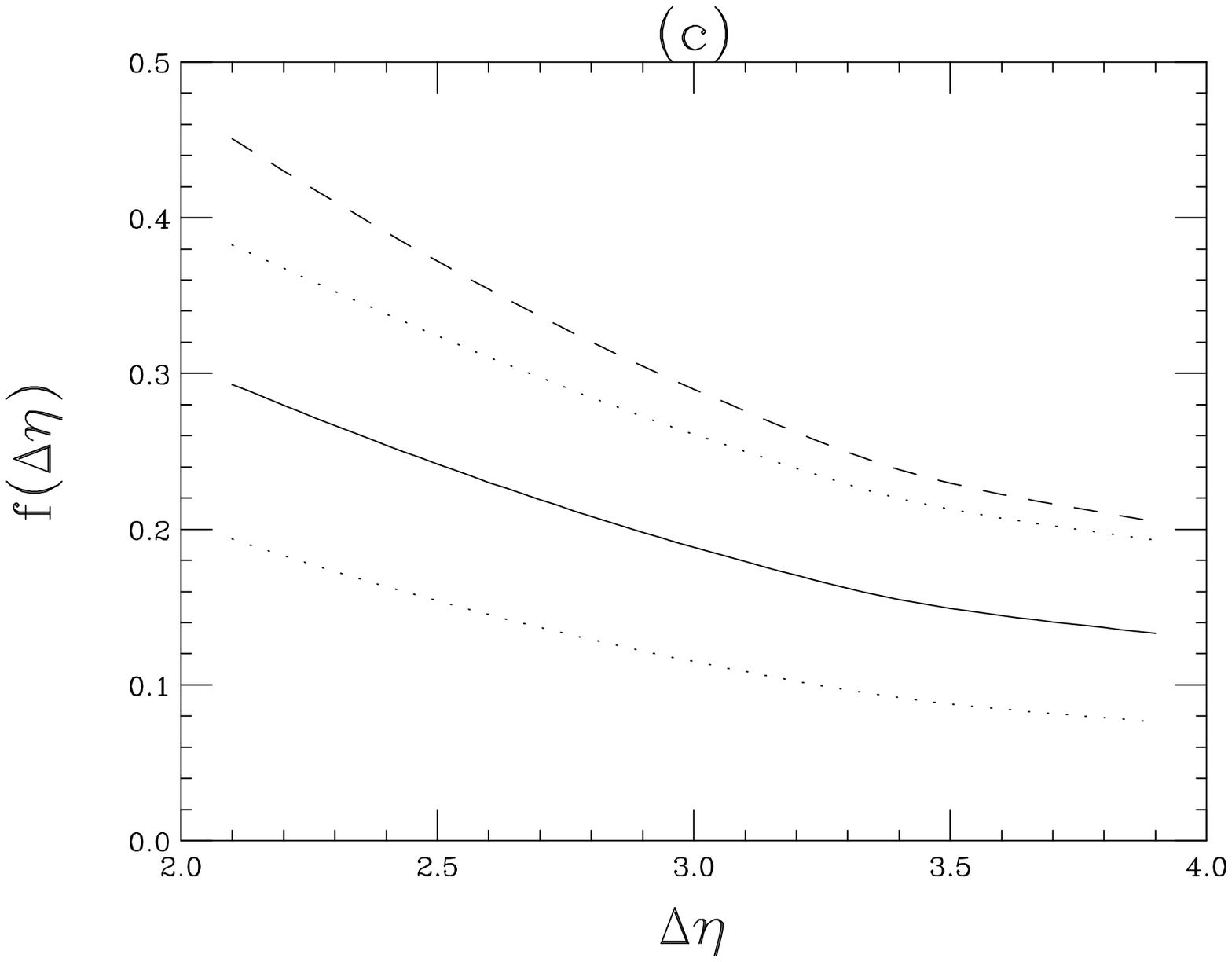}	
\end{minipage}
\hfill
\begin{minipage}{0.48\textwidth}
\includegraphics*[width=7cm,height=6cm]{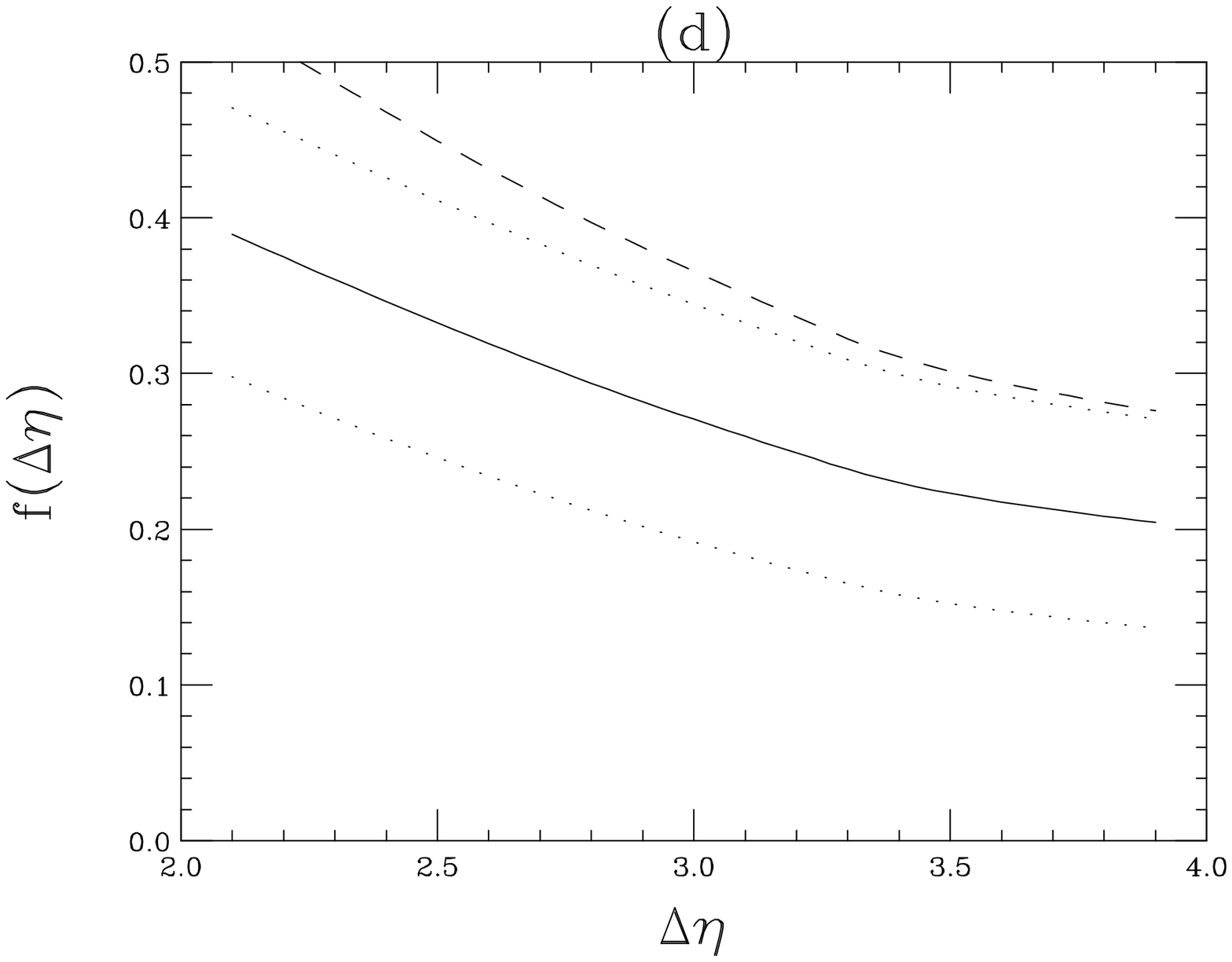}
\end{minipage}\\
\caption{The gap fractions for the ZEUS analysis with a kt defined final state ($R=1.0$), 
at varying $Q_{\Omega}$. $Q_{\Omega}=0.5,\,1.0,\,1.5,\,2.0$~GeV for plots (a), (b), (c) and (d) respectively.
The solid line includes the effects of primary emission and the secondary emission suppression factor. 
The overall theoretical
uncertainty, including the primary uncertainty and the secondary uncertainty, is shown by the
dotted lines. The dashed line indicates the gap fraction obtained by including only primary emission.}
\label{figz2frac}}
The left hand side of figure \ref{figgap} shows the totally inclusive dijet cross section for the H1 analysis and the 
gap cross section
 for $Q_{\Omega}=1.0$~GeV. We have not shown further values of $Q_{\Omega}$ as all the plots show qualitatively the same
behaviour. 
We have cross-checked our total inclusive cross section against the Monte Carlo event generator HERWIG 
\cite{Corcella:2000bw,Corcella:2002jc} and we 
obtained complete agreement for the H1 and both the ZEUS sets of cuts. 
In figure \ref{figgap} the solid curve is
the total inclusive cross section, the dashed line is the cross section with the primary interjet logarithms resummed 
and the
dotted lines show the theoretical uncertainty of the primary resummation, estimated by varying $\alpha_s$ as described 
above. The inclusion of the primary gap
logarithms gives a substantial suppression of the cross section; our analysis confirms simple ``area of phase space''
arguments which say that the kt defined final state will have greater soft gluon suppression than a cone defined
final state due to the increased gap area in the $(\eta,\phi)$ plane. This plot for the ZEUS analysis is shown 
in the right hand side of figure \ref{figgap}.

\subsection{Gap fractions}

The gap fraction is defined as the gap cross section, at fixed $Q_{\Omega}$, divided by the total inclusive cross section.
Figure \ref{figh1frac} shows the gap fraction for the H1 cuts at the four experimentally measured values of 
$Q_{\Omega}$ and figure \ref{figz2frac} shows the gap fractions for the ZEUS analysis. 
The solid line is the gap fraction curve obtained by including the primary emission and the 
NG suppression factors of table~\ref{nglsupp} in the prediction for the gap cross section. The dotted lines show
the theoretical uncertainty of both the primary and secondary emission probabilities, and the dashed line shows the 
gap fraction obtained by including only the primary emission contribution.
We find that our gap fraction is consistent with the H1 values for the
measured $Q_{\Omega}$. The large uncertainty in the gap fraction predictions comes from an approximate 
treatment of the NG suppression and from using perturbation theory at $\sim$1 GeV. Nonetheless, this uncertainty is 
principally in the normalisation of the curves and we expect our resummation to accurately describe 
the shape of the gap fraction curves. 

\section{Conclusions}

\label{secconc}

In this paper we have computed resummed predictions for rapidity gap processes at HERA. We include primary logarithms 
using the soft gluon techniques of CSS, and include the effects of NGLs using an overall 
suppression factor computed from an extension of our earlier work. The kt definition of a hadronic final state 
determines the
phase space available for soft primary emission and we have computed a set of anomalous dimension matrices specific 
to the geometry of the H1 and ZEUS analyses. Of course this method can be used for any definition of the gap, 
provided $\Omega$ is directed away
from all hard jets. 
We then compared our predictions with gaps-between-jets data from the H1 collaboration and found a
consistent agreement. The theoretical uncertainty of our predictions is relatively large, and 
generally dominated by the secondary emission
uncertainty. However our resummed predictions correctly predict the shape of the H1 data, and the normalisation 
agrees within errors. There is a suggestion that the $Q_{\Omega}$ dependence is not quite right, with our central 
$Q_{\Omega}=0.5$~GeV prediction below data and our central $Q_{\Omega}=2.0$~GeV prediction above data, although all 
are within our uncertainty. It is possible that a more complete treatment of the perturbative/non-perturbative interface 
would improve this. We expect that calculation of primary emission will be correct if $Q_{\Omega}$ is not too large, so that we
can neglect jet recoil. However our
 calculation is of sufficient accuracy in the region of phase space probed at HERA.

Our treatment of the NGLs is very approximate. For a fuller treatment, it is necessary to extend the 
extraction of the suppression factor to beyond the large $N_c$ limit and overcome the inherent disadvantages of
the numerical methods used. For the current application, consideration of the four jet system is also necessary. We reserve 
the latter extension, in the large $N_c$ limit, for future work.
Our calculation has not included power corrections \cite{Korchemsky:1999kt}. The inclusion of such 
non-perturbative effects is required for a full and correct comparison to the experimental data. Again, we reserve 
this for future work.

Our calculation involves a numerical integration over all kinematic variables, so it would be straightforward 
to calculate the dependence of the gap fraction on, for example, the fraction of the photon's momentum 
participating in the hard process,~$x_{\gamma}$. This code is available from the authors.

In conclusion, we have shown that the calculation of primary and secondary emission patterns can give a good description 
of rapidity gap data at HERA. A fuller treatment would refine our approximation of NGLs and include power corrections.

\section*{Acknowledgements}
We are grateful to George Sterman and Carola Berger for useful communications and 
RBA would like to acknowledge the Particle Physics and Astronomy Research Council for
financial support.

\begin{appendix}

\section{Colour bases}

\label{appbases}

In this section we present the colour bases used in this work. All the bases in this section have appeared
elsewhere \cite{Oderda:1999kr,Berger:2001ns,Kidonakis:2000gi,Berger:2003zh}, but we show them here for completeness.

\subsection*{The process \boldmath{$q\bar{q}\rightarrow q\bar{q}$}}
\begin{eqnarray}
c_1&=&\delta_{a1}\delta_{b2}, \nonumber \\
c_2&=&-\frac{1}{2N_c}\delta_{a1}\delta_{b2}+\frac12\delta_{ab}\delta_{12}.
\label{qqbarbasis}
\end{eqnarray}

\subsection*{The process \boldmath{$qq\rightarrow qq$}}
\begin{eqnarray}
c_1&=&\delta_{a1}\delta_{b2}, \nonumber \\
c_2&=&-\frac{1}{2N_c}\delta_{a1}\delta_{b2}+\frac12\delta_{a2}\delta_{b1}.
\label{qqbasis}
\end{eqnarray}

\subsection*{The process \boldmath{$qg\rightarrow qg$}}
\begin{eqnarray}
c_1&=&\delta_{a1}\delta_{b2}, \nonumber \\
c_2&=&d_{b2c}(T_F^c)_{1a}, \nonumber \\
c_3&=&if_{b2c}(T^c_F)_{1a}.
\label{qgbasis}
\end{eqnarray}

\subsection*{The processes \boldmath{$q\bar{q}\rightarrow gg$} and \boldmath{$gg\rightarrow q\bar{q}$}}
The process $gg\rightarrow q\bar{q}$ has the basis,
\begin{eqnarray}
c_1&=&\delta_{ab}\delta_{12}, \nonumber \\
c_2&=&d_{abc}(T_F^c)_{12}, \nonumber \\
c_3&=&if_{abc}(T_F^c)_{12}.
\label{qqbarggbasis}
\end{eqnarray}
To find the basis for $q\bar{q}\rightarrow gg$, we interchange $a\leftrightarrow 2$ and $b\leftrightarrow 1$.

\subsection*{The process \boldmath{$gg\rightarrow gg$}}
The complete basis is
\begin{equation}
\left\{
c_1,c_2,c_3,P_1,P_{8_S},P_{8_A},P_{10\bigoplus \overline{10}},P_{27}\right\},
\label{ggbasis}
\end{equation}
where
\begin{eqnarray}
c_1&=&\frac{i}{4}\left[f_{abc}d_{12c}-d_{abc}f_{12c}\right], \nonumber \\
c_2&=&\frac{i}{4}\left[f_{abc}d_{12c}+d_{abc}f_{12c}\right], \nonumber \\
c_3&=&\frac{i}{4}\left[f_{a1c}d_{b2c}+d_{a1c}f_{b2c}\right], \nonumber \\
P_1&=&\frac{1}{8}\delta_{a1}\delta_{b2}, \nonumber \\
P_{8_S}&=&\frac{3}{5}d_{a1c}d_{b2c}, \nonumber \\
P_{8_A}&=&\frac{1}{3}f_{a1c}f_{b2c}, \nonumber \\
P_{10\bigoplus\overline{10}}&=&\frac12\left(\delta_{ab}\delta_{12}-\delta_{a2}\delta_{b1}\right)-
\frac13 f_{a1c}f_{b2c}, \nonumber \\
P_{27}&=&\frac12 \left(\delta_{ab}\delta_{12}+\delta_{a2}\delta_{b1}\right)-\frac18\delta_{a1}\delta_{b2}
-\frac35 d_{a1c} d_{b2c}.
\end{eqnarray}

\subsection*{The direct processes}

Since there is only one colour structure, these are basis independent.

\section{The hard and soft matrices}

\label{apphardsoft}

We now show the complete set of hard and soft matrices used in this work. These matrices have appeared 
in a variety of forms in \cite{Oderda:1999kr,Berger:2001ns,Kidonakis:2000gi,Berger:2003zh}. In all these equations we
have set $N_c=3$ and have written the coupling scale as $\mu$. Note that all our hard matrices differ from 
the normalisation used in~\cite{Oderda:1999kr,Kidonakis:2000gi} by 
a factor of~$\pi/(2\hat{s})\,4\hat{t}\hat{u}/\hat{s}^2$, while
they agree with that used in~\cite{Berger:2001ns}.

\subsection*{The process \boldmath{$q\bar{q}\rightarrow q\bar{q}$}}

The hard matrix has, in the basis \ref{qqbarbasis}, the form
\begin{equation}
H^{(1)}=\frac{1}{9}\frac{\alpha_s^2(\mu)\pi}{\hat{s}}
\;\frac{4\hat{t}\hat{u}}{\hat{s}^2}
\left(\begin{array}[c]{cc}
\frac{16}{81} \chi_1 & \frac{4}{27}\chi_2 \\
\frac{4}{27} \chi_2 & \chi_3 
\end{array}\right),
\end{equation}
where we define
\begin{eqnarray}
\chi_1&=&\frac{\hat{t}^2+\hat{u}^2}{\hat{s}^2}, \nonumber \\
\chi_2&=&3\frac{\hat{u}^2}{\hat{s}\hat{t}}-\frac{\hat{t}^2+\hat{u}^2}{\hat{s}^2}, \nonumber \\
\chi_3&=&\frac{\hat{s}^2+\hat{u}^2}{\hat{t}^2}+\frac19\frac{\hat{t}^2+\hat{u}^2}{\hat{s}^2}-\frac23\frac{\hat{u}^2}
{\hat{s}\hat{t}}.
\end{eqnarray}
The unequal flavour process $q\bar{q}'\rightarrow q\bar{q}'$ is found by dropping the $s$-channel terms from 
these equations, and the unequal flavour process $q\bar{q}\rightarrow q'\bar{q}'$ is found by dropping the
$t$-channel terms. The hard matrix for $q\bar{q}\rightarrow \bar{q}q$ is found using the
transformation $\hat{t} \leftrightarrow \hat{u}$.
The corresponding soft matrix for all these processes is
\begin{equation}
S^{(0)}=
\left(\begin{array}[c]{cc}
N_c^2 & 0 \\
0 & \frac{1}{4}(N_c^2-1)
\end{array}\right).
\end{equation}

\subsection*{The process \boldmath{$qq\rightarrow qq$}}

The hard matrix has, in the basis \ref{qqbasis}, the form
\begin{equation}
H^{(1)}=\frac{1}{9}\frac{\alpha_s^2(\mu)\pi}{\hat{s}}
\;\frac{4\hat{t}\hat{u}}{\hat{s}^2}
\left(\begin{array}[c]{cc}
\frac{16}{81} \chi_1 & \frac{4}{27}\chi_2 \\
\frac{4}{27} \chi_2 & \chi_3 
\end{array}\right),
\end{equation}
where we define
\begin{eqnarray}
\chi_1&=&\frac{\hat{t}^2+\hat{s}^2}{\hat{u}^2}, \nonumber \\
\chi_2&=&3\frac{\hat{s}^2}{\hat{u}\hat{t}}-\frac{\hat{t}^2+\hat{s}^2}{\hat{u}^2}, \nonumber \\
\chi_3&=&\frac{\hat{u}^2+\hat{s}^2}{\hat{t}^2}+\frac19\frac{\hat{t}^2+\hat{s}^2}{\hat{u}^2}-\frac23\frac{\hat{s}^2}
{\hat{u}\hat{t}}.
\end{eqnarray}
For the process $qq'\rightarrow qq'$ only keep the $t$-channel terms.
The corresponding soft matrix is
\begin{equation}
S^{(0)}=
\left(\begin{array}[c]{cc}
N_c^2 & 0 \\
0 & \frac{1}{4}(N_c^2-1)
\end{array}\right).
\end{equation}

\subsection*{The process \boldmath{$qg\rightarrow qg$}}

The hard matrix has, in the basis \ref{qgbasis}, the form
\begin{equation}
H^{(1)}=\frac{1}{24}\frac{\alpha_s^2(\mu)\pi}{2\hat{s}}
\;\frac{4\hat{t}\hat{u}}{\hat{s}^2}
\left(\begin{array}[c]{ccc}
\frac{1}{18} \chi_1 & \frac{1}{6}\chi_1 & \frac13\chi_2 \\
\frac{1}{6} \chi_1 & \frac12\chi_1 & \chi_2 \\
\frac13\chi_2 &  \chi_2 & \chi_3
\end{array}\right),
\end{equation}
where we define
\begin{eqnarray}
\chi_1&=&2-\frac{\hat{t}^2}{\hat{s}\hat{u}}, \nonumber \\
\chi_2&=&1-\frac12 \frac{\hat{t}^2}{\hat{s}\hat{u}}-\frac{\hat{u}^2}{\hat{s}\hat{t}}-\frac{\hat{s}}{\hat{t}}, \nonumber \\
\chi_3&=&3-4\frac{\hat{s}\hat{u}}{\hat{t}^2}-\frac12 \frac{\hat{t}^2}{\hat{s}\hat{u}}.
\end{eqnarray}
The hard matrix for the process $qg\rightarrow gq$ is found by the transformation $\hat{t}\leftrightarrow \hat{u}$.
The corresponding soft matrix is
\begin{equation}
S^{(0)}=
\left(\begin{array}[c]{ccc}
N_c(N_c^2-1) & 0 & 0\\
0 & \frac{1}{2N_c}(N_c^2-4)(N_c^2-1) & 0 \\
0 & 0 & \frac12 N_c (N_c^2-1)
\end{array}\right).
\end{equation}

\subsection*{The processes \boldmath{$q\bar{q}\rightarrow gg$} and \boldmath{$gg\rightarrow q\bar{q}$}}
In the basis \ref{qqbarggbasis} the hard matrix for these processes has the form
\begin{equation}
H^{(1)}=\frac{1}{\Delta}\frac{\alpha_s^2(\mu)\pi}{2\hat{s}}
\;\frac{4\hat{t}\hat{u}}{\hat{s}^2}
\left(\begin{array}[c]{ccc}
\frac{1}{18} \chi_1 & \frac{1}{6}\chi_1 & \frac16\chi_2 \\
\frac{1}{6} \chi_1 & \frac12\chi_1 & \frac12\chi_2 \\
\frac16\chi_2 &  \frac12\chi_2 & \frac12\chi_3
\end{array}\right),
\end{equation}
where we define
\begin{eqnarray}
\chi_1&=&\frac{\hat{t}^2+\hat{u}^2}{\hat{t}\hat{u}}, \nonumber \\
\chi_2&=&\left(1+\frac{2\hat{t}}{\hat{s}}\right)\chi_1, \nonumber \\
\chi_3&=&\left(1-\frac{4\hat{t}\hat{u}}{\hat{s}^2}\right)\chi_1.
\end{eqnarray}
The constant $\Delta=9$ for the process $q\bar{q}\rightarrow gg$ and 
$\Delta=64$ for the process $gg\rightarrow q\bar{q}$. The matrix for the process 
$gg\rightarrow \bar{q}q$ is found from the transformation $\hat{t}\leftrightarrow \hat{u}$.
The soft matrix is
\begin{equation}
S^{(0)}=\frac{N_c^2-1}{2N_c}
\left(\begin{array}[c]{ccc}
2N_c^2 & 0 & 0 \\
0 & N_c^2-4 & 0 \\
0 & 0 & N_c^2 
\end{array}\right).
\end{equation}

\subsection*{The process \boldmath{$gg \rightarrow gg$}}
The hard matrix, in the basis \ref{ggbasis} has the block-diagonal form
\begin{equation}
H^{(1)}=
\left(\begin{array}[c]{cc}
0_{3\times3} & 0_{3\times5} \\
0_{5\times3} & H^{(1)}_{5\times5} 
\end{array}\right),
\end{equation}
where the matrix $H^{(1)}_{5\times5}$ has the form
\begin{equation}
H^{(1)}_{5\times5}=\frac{1}{16}\frac{\alpha_s^2(\mu)\pi}{2\hat{s}}
\;\frac{4\hat{t}\hat{u}}{\hat{s}^2}
\left(\begin{array}[c]{ccccc}
9\chi_1 & \frac92\chi_1 & \frac92\chi_2 & 0 & -3\chi_1 \\
\frac92\chi_1 & \frac94\chi_1 & \frac94 \chi_2 &  0 & -\frac32\chi_1 \\
\frac92\chi_2 & \frac94\chi_2 & \chi_3 & 0 & -\frac32\chi_2 \\
0 & 0 & 0 & 0 & 0 \\
-3\chi_1 & -\frac32\chi_1 & -\frac32\chi_2 & 0 & \chi_1
\end{array}\right),
\end{equation}
and we write
\begin{eqnarray}
\chi_1&=&1-\frac{\hat{t}\hat{u}}{\hat{s}^2}-\frac{\hat{s}\hat{t}}{\hat{u}^2}+\frac{\hat{t}^2}{\hat{s}\hat{u}}, \nonumber \\
\chi_2&=&\frac{\hat{s}\hat{t}}{\hat{u}^2}-\frac{\hat{t}\hat{u}}{\hat{s}^2}+\frac{\hat{u}^2}{\hat{s}\hat{t}}
-\frac{\hat{s}^2}{\hat{t}\hat{u}}, \nonumber \\
\chi_3&=&\frac{27}{4}
-9\left(\frac{\hat{s}\hat{u}}{\hat{t}^2}+\frac14\frac{\hat{t}\hat{u}}{\hat{s}^2}+
\frac14\frac{\hat{s}\hat{t}}{\hat{u}^2}\right)
+\frac92\left(\frac{\hat{u}^2}{\hat{s}\hat{t}}+\frac{\hat{s}^2}{\hat{t}\hat{u}}-
\frac12\frac{\hat{t}^2}{\hat{s}\hat{u}}\right).
\end{eqnarray}
For this process the soft matrix is
\begin{equation}
S^{(0)}=
\left(\begin{array}[c]{cccccccc}
5 & 0 & 0 & 0 & 0 & 0 & 0 & 0 \\
0 & 5 & 0 & 0 & 0 & 0 & 0 & 0 \\
0 & 0 & 5 & 0 & 0 & 0 & 0 & 0 \\
0 & 0 & 0 & 1 & 0 & 0 & 0 & 0 \\
0 & 0 & 0 & 0 & 8 & 0 & 0 & 0 \\
0 & 0 & 0 & 0 & 0 & 8 & 0 & 0 \\
0 & 0 & 0 & 0 & 0 & 0 & 20 & 0 \\
0 & 0 & 0 & 0 & 0 & 0 & 0 & 27 
\end{array}\right).
\end{equation}

\subsection*{The direct processes}
For both these processes the zeroth order soft factor is unity and the 
hard functions are
\begin{eqnarray}
H^{(1)}(\gamma g \rightarrow q\bar{q})&=&
\Bigl(\sum_qe_q^2\Bigr)
\frac{\alpha_s \alpha_{\mathrm{em}}\pi}{2\hat{s}}
\;\frac{4\hat{t}\hat{u}}{\hat{s}^2}
\left(\frac{\hat{u}}{\hat{t}}+\frac{\hat{t}}{\hat{u}}\right), \nonumber \\
H^{(1)}(\gamma q(\bar{q}) \rightarrow g q(\bar{q}))&=&\frac83e_q^2\frac{\alpha_s\alpha_{\mathrm{em}}\pi}{2\hat{s}}
\;\frac{4\hat{t}\hat{u}}{\hat{s}^2}
\left(\frac{-\hat{u}}{\hat{s}}+\frac{\hat{s}}{-\hat{u}}\right),
\end{eqnarray}
where $e_q$ is the electric charge of quark flavour $q$, in units of the electron charge.  Note that if the
sum for $\gamma g \rightarrow q\bar{q}$ is taken to be over four flavours, then this gives a factor of~$10/9$.

\section{Colour decomposition matrices}

\label{appdecomp}

We now give the full set of colour decomposition matrices, and the 
sign function $\mathcal{S}$, defined by equation \ref{eqsign}, for $\alpha$, $\beta$ and $\gamma$, defined 
by 
\begin{eqnarray}
\alpha&=&\mathcal{S}_{ab}\Gamma^{(ab)}+\mathcal{S}_{12}\Gamma^{(12)}, \nonumber \\
\beta&=&\mathcal{S}_{a1}\Gamma^{(a1)}+\mathcal{S}_{b2}\Gamma^{(b2)}, \nonumber \\
\gamma&=&\mathcal{S}_{b1}\Gamma^{(b1)}+\mathcal{S}_{a2}\Gamma^{(a2)}.
\end{eqnarray}
\subsection*{The process \boldmath{$q\bar{q}\rightarrow q\bar{q}$}}
\begin{equation}
\mathcal{C}^{q\bar{q}\rightarrow q\bar{q}}=
\left(\begin{array}[c]{cc}
C_F \beta \,\,\,& \frac{C_F}{2N_c}(\alpha+\gamma) \\
\alpha+\gamma \,\,\,&  C_F\alpha-\frac{1}{2N_c}(\alpha+\beta+2\gamma)
\end{array}\right).
\end{equation}
The signs are
\begin{eqnarray}
\mathcal{S}_\alpha&=&+1, \\
\mathcal{S}_\beta&=&+1, \\
\mathcal{S}_\gamma&=&-1.
\end{eqnarray}

\subsection*{The process \boldmath{$qq\rightarrow qq$}}
\begin{equation}
\mathcal{C}^{qq\rightarrow qq}=
\left(\begin{array}[c]{cc}
C_F \beta \,\,\,& \frac{C_F}{2N_c}(\alpha+\gamma) \\
\alpha+\gamma \,\,\,&  C_F\gamma-\frac{1}{2N_c}(2\alpha+\beta+\gamma)
\end{array}\right).
\end{equation}
The signs are
\begin{eqnarray}
\mathcal{S}_\alpha&=&-1, \\
\mathcal{S}_\beta&=&+1, \\
\mathcal{S}_\gamma&=&+1.
\end{eqnarray}

\subsection*{The process \boldmath{$qg\rightarrow qg$}}
\begin{equation}
\mathcal{C}^{qg\rightarrow qg}=
\left(\begin{array}[c]{ccc}
C_F \Gamma^{(a1)}+C_A\Gamma^{(b2)} & 0 &  -\frac12(\alpha+\gamma) \\
0 & \chi & -\frac{N_c}{4}(\alpha+\gamma) \\
-(\alpha+\gamma) & -\frac{N_c^2-4}{4N_c}(\alpha+\gamma) & \chi
\end{array}\right).
\end{equation}
The signs are
\begin{eqnarray}
\mathcal{S}_\alpha&=&+1, \\
\mathcal{S}_\beta&=&+1, \\
\mathcal{S}_\gamma&=&-1,
\end{eqnarray}
and we define
\begin{equation}
\chi=\frac{N_c}{4}(\alpha-\gamma)-\frac{1}{2N_c}\Gamma^{(a1)}+\frac{N_c}{2}\Gamma^{(b2)}.
\end{equation}

\subsection*{The processes \boldmath{$q\bar{q}\rightarrow gg$} and \boldmath{$gg \rightarrow q\bar{q}$}}

For $q\bar{q}\rightarrow gg$ we have
\begin{equation}
\mathcal{C}^{q\bar{q}\rightarrow gg}=
\left(\begin{array}[c]{ccc}
C_F \Gamma^{(ab)}+C_A\Gamma^{(12)} & 0 &  \frac12(\beta+\gamma) \\
0 & \chi' & \frac{N_c}{4}(\beta+\gamma) \\
(\beta+\gamma) & \frac{N_c^2-4}{4N_c}(\beta+\gamma) & \chi'
\end{array}\right).
\end{equation}
The signs are
\begin{eqnarray}
\mathcal{S}_\alpha&=&+1, \\
\mathcal{S}_\beta&=&+1, \\
\mathcal{S}_\gamma&=&-1,
\end{eqnarray}
and we define
\begin{equation}
\chi'=\frac{N_c}{4}(\beta-\gamma)-\frac{1}{2N_c}\Gamma^{(ab)}+\frac{N_c}{2}\Gamma^{(12)}.
\end{equation}

For $gg\rightarrow q\bar{q}$ we have
\begin{equation}
\mathcal{C}^{gg\rightarrow q\bar{q}}=
\left(\begin{array}[c]{ccc}
C_F \Gamma^{(12)}+C_A\Gamma^{(ab)} & 0 &  \frac12(\beta+\gamma) \\
0 & \chi'' & \frac{N_c}{4}(\beta+\gamma) \\
(\beta+\gamma) & \frac{N_c^2-4}{4N_c}(\beta+\gamma) & \chi''
\end{array}\right).
\end{equation}
The signs are
\begin{eqnarray}
\mathcal{S}_\alpha&=&+1, \\
\mathcal{S}_\beta&=&+1, \\
\mathcal{S}_\gamma&=&-1,
\end{eqnarray}
and we define
\begin{equation}
\chi''=\frac{N_c}{4}(\beta-\gamma)-\frac{1}{2N_c}\Gamma^{(12)}+\frac{N_c}{2}\Gamma^{(ab)}.
\end{equation}

\subsection*{The process \boldmath{$gg\rightarrow gg$}}
\begin{equation}
\mathcal{C}^{gg\rightarrow gg}=
\left(\begin{array}[c]{cc}
\mathcal{M}_{3\times3} & 0_{3\times5} \\
0_{5\times3} & \mathcal{M}_{5\times5}
\end{array}\right),
\end{equation}
where the matrix $\mathcal{M}_{3\times3}$ is 
\begin{equation}
\mathcal{M}_{3\times3}=
\left(\begin{array}[c]{ccc}
\frac{N_c}{2}(\alpha+\beta) & 0 & 0 \\
0 & \frac{N_c}{2}(\alpha-\gamma) & 0 \\
0 & 0 & \frac{N_c}{2}(\beta-\gamma),
\end{array}\right)
\end{equation}
and the matrix $\mathcal{M}_{5\times5}$ is 
\begin{equation}
\mathcal{M}_{5\times5}=
\left(\begin{array}[c]{ccccc}
3\beta & 0 & 3(\alpha+\gamma) & 0 & 0 \\
0 & \frac{3}{4}(\alpha+2\beta-\gamma) & \frac{3}{4}(\alpha+\gamma) & \frac{3}{2}(\alpha+\gamma) & 0 \\
\frac{3}{8}(\alpha+\gamma)&\frac{3}{4}(\alpha+\gamma)&\frac{3}{4}(\alpha+2\beta-\gamma) & 0 & \frac{9}{8}(\alpha+\gamma) \\
0 & \frac{3}{5}(\alpha+\gamma) & 0 & \frac{3}{2}(\alpha-\gamma) & \frac{9}{10}(\alpha+\gamma) \\
0 & 0 & \frac{1}{3}(\alpha+\gamma) & \frac{2}{3}(\alpha+\gamma) & 2\alpha-\beta-2\gamma
\end{array}\right),
\end{equation}
for $N_c=3$.
The signs are
\begin{eqnarray}
\mathcal{S}_\alpha&=&+1, \\
\mathcal{S}_\beta&=&+1, \\
\mathcal{S}_\gamma&=&-1.
\end{eqnarray}

\subsection*{The direct processes}
This processes has no matrix structure.
\begin{eqnarray}
\mathcal{C}^{\gamma g \rightarrow q\bar{q}}&=&
-\frac{1}{2N_c}\Gamma^{(12)}+\frac{N_c}{2}\left(\Gamma^{(b1)}+\Gamma^{(b2)}\right), \nonumber \\
\mathcal{C}^{\gamma q \rightarrow gq}&=&
-\frac{1}{2N_c}\Gamma^{(b2)}+\frac{N_c}{2}\left(\Gamma^{(b2)}+\Gamma^{(12)}\right).
\end{eqnarray}

\section{The \boldmath{$\Gamma^{(ij)}$} series expansions}

\label{appgamma}

We have not found a closed form for these integrals, but they are
straightforward to express as power series in $R$ and $e^{-\Delta\eta}$
(by Lorentz invariance, only the contributions from dipoles containing
jet 2 are $\Delta\eta$-dependent),
\begin{eqnarray}
\Omega^{(ab)}_1 &=& \frac{\alpha_s}{\pi}\Bigl(
\frac14R^2\Bigr), \\
\Omega^{(12)}_1 &=& \frac{\alpha_s}{\pi}\Bigl(
\frac12\log R+\frac12\log\frac1{\Delta\eta-\Delta y}+
\\&&
(+ 0.31831\,R + 0.06250\,R^2 + 0.00884\,R^3 + 0.00087\,R^4 +
   0.00003\,R^5)+
\hspace*{-1cm}\nonumber\\&&
(- 0.08616\,R - 0.03383\,R^2 -
   0.01197\,R^3 - 0.00282\,R^4
\nonumber\\&&
 - 0.00039\,R^5 +
   0.00001\,R^7)e^{-(\Delta\eta-2)}+
\nonumber\\&&
(+ 0.01166\,R + 0.00916\,R^2 +
   0.00551\,R^3 + 0.00305\,R^4
\nonumber\\&&
 + 0.00122\,R^5 +
   0.00038\,R^6 + 0.00011\,R^7 + 0.00003\,R^8)e^{-2(\Delta\eta-2)}+
\nonumber\\&&
(- 0.00158\,R - 0.00186\,R^2 - 0.00162\,R^3 -
   0.00139\,R^4
\nonumber\\&&
 - 0.00088\,R^5 - 0.00041\,R^6 -
   0.00017\,R^7 - 0.00007\,R^8 - 0.00002\,R^9)e^{-3(\Delta\eta-2)}+
\hspace*{-3cm}\nonumber\\&&
(+ 0.00021\,R +
   0.00034\,R^2 + 0.00039\,R^3 + 0.00045\,R^4
\nonumber\\&&
 + 0.00039\,R^5 + 0.00024\,R^6 + 0.00013\,R^7 +
   0.00007\,R^8
\nonumber\\&&
 + 0.00003\,R^9 +
   0.00001\,R^{10})e^{-4(\Delta\eta-2)}+
\nonumber\\&&
(- 0.00003\,R -
   0.00006\,R^2 - 0.00008\,R^3 - 0.00012\,R^4
\nonumber\\&&
 - 0.00013\,R^5 - 0.00010\,R^6 - 0.00007\,R^7 -
   0.00004\,R^8
\nonumber\\&&
 - 0.00003\,R^9 -
   0.00001\,R^{10})e^{-5(\Delta\eta-2)}+
\nonumber\\&&
(+ 0.00001\,R^2 + 0.00002\,R^3 +
   0.00003\,R^4 + 0.00004\,R^5
\nonumber\\&&
 + 0.00003\,R^6 +
   0.00003\,R^7 + 0.00002\,R^8 + 0.00001\,R^9)e^{-6(\Delta\eta-2)}
\Bigr),
\nonumber\\
\Omega^{(a1)}_1 &=& \frac{\alpha_s}{\pi}\Bigl(
\frac12\log R+\frac12\log\frac1{\Delta\eta-\Delta y}
\\&&{}
 - 0.31831\,R + 0.06250\,R^2 - 0.00884\,R^3 + 0.00087\,R^4 -
   0.00003\,R^5
\Bigr),
\hspace*{-1cm}
\nonumber\\
\Omega^{(b2)}_1 &=& \frac{\alpha_s}{\pi}\Bigl(
\\&&{}
(+ 0.00458\,R^2 + 0.00389\,R^3 + 0.00229\,R^4 +
   0.00104\,R^5
\nonumber\\&&
 + 0.00038\,R^6 + 0.00012\,R^7 +
   0.00003\,R^8)e^{-2(\Delta\eta-2)}+
\nonumber\\&&
(- 0.00124\,R^2 -
   0.00158\,R^3 - 0.00124\,R^4 - 0.00079\,R^5
\nonumber\\&&
 - 0.00041\,R^6 - 0.00018\,R^7 - 0.00007\,R^8 -
   0.00002\,R^9)e^{-3(\Delta\eta-2)}+
\nonumber\\&&
(+ 0.00025\,R^2 + 0.00043\,R^3 + 0.00042\,R^4 +
   0.00034\,R^5
\nonumber\\&&
 + 0.00024\,R^6 + 0.00014\,R^7 +
   0.00007\,R^8 + 0.00003\,R^9 +
   0.00001\,R^{10})e^{-4(\Delta\eta-2)}+
\hspace*{-3cm}\nonumber\\&&
(- 0.00005\,R^2 -
   0.00010\,R^3 - 0.00011\,R^4 - 0.00011\,R^5
\nonumber\\&&
 - 0.00010\,R^6 - 0.00007\,R^7 - 0.00004\,R^8 -
   0.00002\,R^9 - 0.00001\,R^{10})e^{-5(\Delta\eta-2)}+
\hspace*{-3cm}\nonumber\\&&
(+ 0.00002\,R^3 +
   0.00003\,R^4 + 0.00003\,R^5 + 0.00003\,R^6
\nonumber\\&&
 + 0.00003\,R^7 + 0.00002\,R^8 + 0.00001\,R^9)e^{-6(\Delta\eta-2)}
\Bigr),
\nonumber\\
\Omega^{(a2)}_1 &=& \frac{\alpha_s}{\pi}\Bigl(
-\frac14R^2+
\\&&{}
(+ 0.06767\,R^2 + 0.02872\,R^3 - 0.00096\,R^5 +
   0.00002\,R^7)e^{-(\Delta\eta-2)}+
\nonumber\\&&
(- 0.01374\,R^2 -
   0.01166\,R^3 - 0.00229\,R^4 - 0.00038\,R^6
\nonumber\\&&
 - 0.00021\,R^7 - 0.00003\,R^8)e^{-2(\Delta\eta-2)}+
\nonumber\\&&
(+ 0.00248\,R^2 + 0.00316\,R^3 + 0.00124\,R^4 +
   0.00032\,R^5
\nonumber\\&&
 + 0.00041\,R^6 + 0.00027\,R^7 +
   0.00007\,R^8 + 0.00001\,R^9)e^{-3(\Delta\eta-2)}+
\nonumber\\&&
(- 0.00042\,R^2 -
   0.00071\,R^3 - 0.00042\,R^4 - 0.00019\,R^5
\nonumber\\&&
 - 0.00024\,R^6 - 0.00019\,R^7 - 0.00007\,R^8 -
   0.00002\,R^9 - 0.00001\,R^{10})e^{-4(\Delta\eta-2)}+
\hspace*{-3cm}\nonumber\\&&
(+ 0.00007\,R^2 + 0.00014\,R^3 + 0.00011\,R^4 +
   0.00007\,R^5
\nonumber\\&&
 + 0.00010\,R^6 + 0.00009\,R^7 +
   0.00004\,R^8 + 0.00002\,R^9 +
   0.00001\,R^{10})e^{-5(\Delta\eta-2)}+
\hspace*{-3cm}\nonumber\\&&
(- 0.00001\,R^2 -
   0.00003\,R^3 - 0.00003\,R^4 - 0.00002\,R^5
\nonumber\\&&
 - 0.00003\,R^6 - 0.00004\,R^7 - 0.00002\,R^8 -
   0.00001\,R^9)e^{-6(\Delta\eta-2)}
\Bigr),
\nonumber\\
\Omega^{(b1)}_1 &=& \frac{\alpha_s}{\pi}\Bigl(
-\frac12\log R-\frac12\log\frac1{\Delta\eta-\Delta y}
\\&&{}
 - 0.31831\,R - 0.06250\,R^2 - 0.00884\,R^3 - 0.00087\,R^4 -
   0.00003\,R^5
\Bigr),
\hspace*{-1cm}
\nonumber
\end{eqnarray}
where all coefficients larger than $10^{-5}$ are shown (recall that we
are mainly interested in the case $R=1$, $\Delta\eta>2$). By symmetry, we have $\Omega^{(ij)}_2 =
\Omega^{(\bar\imath\bar\jmath)}_1$, where the mapping $i\to\bar\imath$
is given by $\{a,b,1,2\}\to\{b,a,2,1\}$.

\section{The \boldmath{$\Omega_f^{(ij)}$} angular integrals for a cone geometry}
\label{appcone}
We present these results as they have not appeared previously in this form.
They have the expression,
\begin{equation}
\Omega^{(ij)}_f=\int_{-\Delta y/2}^{+\Delta_y/2}d\eta \int_0^{2\pi} \frac{d\phi}{2\pi} 
\frac{\beta_i\cdot \beta_j}{(\beta_i\cdot \bar{k})(\beta_j\cdot \bar{k})},
\end{equation}
where the integrand is found from the appropriate 4-momenta, and the phase space is taken to
be of width $\Delta y$ and azimuthally symmetric. Note that these expressions do not include
the sign factors. We obtain
\begin{eqnarray}
\Omega_f^{(ab)}&=&2\Delta y, \nonumber \\
\Omega_f^{(12)}&=&2\log\left(\frac{\sinh(\Delta\eta/2+\Delta y/2)}{\sinh(\Delta\eta/2-\Delta y/2)}\right), \nonumber \\ 
\Omega_f^{(a1)}&=&-\Delta y+
\log\left(\frac{\sinh(\Delta\eta/2+\Delta y/2)}{\sinh(\Delta\eta/2-\Delta y/2)}\right), \nonumber \\ 
\Omega_f^{(b2)}&=&-\Delta y+
\log\left(\frac{\sinh(\Delta\eta/2+\Delta y/2)}{\sinh(\Delta\eta/2-\Delta y/2)}\right), \nonumber \\ 
\Omega_f^{(a2)}&=&\Delta y+
\log\left(\frac{\sinh(\Delta\eta/2+\Delta y/2)}{\sinh(\Delta\eta/2-\Delta y/2)}\right), \nonumber \\ 
\Omega_f^{(b1)}&=&\Delta y+
\log\left(\frac{\sinh(\Delta\eta/2+\Delta y/2)}{\sinh(\Delta\eta/2-\Delta y/2)}\right).
\end{eqnarray}

\end{appendix}

\end{document}